\documentclass[amsmath,10pt,twocolumn,superscriptaddress,footinbib,pre]{revtex4-1}
\usepackage{amsmath,amsfonts,bm,dsfont}
\usepackage{graphicx}
\usepackage{subfigure}
\usepackage{float}
\usepackage{siunitx}
\usepackage{uri}
\usepackage[dvipsnames,usenames]{xcolor} % xcolor must be loaded before tikz to prevent a package clash

\usepackage{color}
\usepackage[colorlinks=true,allcolors=blue]{hyperref}
\usepackage{tikz}

\begin{document}
\title{Computing time-periodic steady-state currents via the time evolution of tensor network states}
\author{Nils E.~Strand}
\author{Hadrien Vroylandt}
\author{Todd R.~Gingrich}
\email{todd.gingrich@northwestern.edu}
\affiliation{Department of Chemistry, Northwestern University, 2145 Sheridan Road, Evanston, Illinois 60208, USA}

\begin{abstract}
We present an approach based upon binary tree tensor network (BTTN) states for computing steady-state current statistics for a many-particle 1D ratchet subject to volume exclusion interactions.
The ratcheted particles, which move on a lattice with periodic boundary conditions subject to a time-periodic drive, can be stochastically evolved in time to sample representative trajectories via a Gillespie method.
In lieu of generating realizations of trajectories, a BTTN state can variationally approximate a distribution over the vast number of many-body configurations.
We apply the density matrix renormalization group (DMRG) algorithm to initialize BTTN states, which are then propagated in time via the time-dependent variational principle (TDVP) algorithm to yield the steady-state behavior, including the effects of both typical and rare trajectories.
The application of the methods to ratchet currents is highlighted in a companion letter, but the approach extends naturally to other interacting lattice models with time-dependent driving.
Though trajectory sampling is conceptually and computationally simpler, we discuss situations for which the BTTN TDVP strategy could be more favorable.
\end{abstract}
\maketitle
\section{Introduction}
Over the past few decades, tensor networks have emerged as one of the most powerful mathematical tools for numerically manipulating high-dimensional quantum states.
The computational power is achieved because tensor networks dramatically shrink the dimensionality of quantum states and operators by decomposing intractably large vectors and matrices into a composition of smaller tensors, offering an attractive approximation method.
Furthermore, by adjusting the dimensionality of auxiliary indices that link the tensors, tensor networks enable variational calculations with controllable errors.
In the limit of high-dimensional auxiliary indices, exact results are recovered, but the practical benefit is gained by reducing the dimensionality to obtain approximate results at a dramatically lower computational expense.
For example, the density matrix renormalization group (DMRG) algorithm~\cite{white1992density} is widely used to converge low-entanglement many-body quantum ground states by sweeping through a tensor network while performing computationally tractable local optimizations.
Dynamics of quantum states can similarly be approximated via the time-dependent variational principle (TDVP)~\cite{dirac1930exchange, frenkel1934wave}, which also proceeds by a sweep of local operations on tractable tensors~\cite{haegeman2011time, haegeman2016unifying}.

While tensor networks were initially applied to quantum systems, their use has also been extended to classical stochastic systems~\cite{helms2019dynamical,banuls2019using,helms2020dynamical,Nagy2002,Hieida1998,Temme2010,Gorissen2009,Johnson2010,Johnson2015}.
More specifically, tensor network methods have been used to compute large deviation functions, which measure the probability of dynamical fluctuations both near equilibrium and far from equilibrium.
Helms et al.\ recently identified dynamical phase transitions separating jamming and flowing phases within the 1D and 2D asymmetric simple exclusion processes (ASEP).
These studies used the matrix product state (MPS), a 1D chain of tensors, and the projected entangled pair state (PEPS), the 2D analog of the MPS, to probe the thermodynamic limits of the ASEP for 1D and 2D systems, respectively~\cite{helms2019dynamical, helms2020dynamical}.
In another study by Ba{\~n}uls et al., DMRG was used to compute currents and trajectory-space phase transitions within kinetically-constrained models~\cite{banuls2019using}.
Still, the use of tensor networks to evaluate classical stochastic dynamics remains relatively unexplored.
A major complication is that the relevant stochastic operators, unlike quantum operators, are almost always non-Hermitian and, compared to Hermitian operators, diagonalizing those non-Hermitian operators is more demanding and more prone to numerical instabilities.
Consequently, iterative tensor network procedures like DMRG can present numerical complications when applied to non-symmetric operators~\cite{carlon1999density}.

Here we show that classical stochastic dynamics can nevertheless by robustly propagated by non-Hermitian operators via TDVP using a binary tree tensor network (BTTN)~\cite{kohn2020superfluid,bauernfeind2020time,Felser2021}.
As discussed in a companion letter, the TDVP approach offers an unexplored route to analyze the impact of many-particle interactions in noise-driven ratchets~\cite{strand2021companion}.
We had previously analyzed a single-particle 2D ratchet's behavior under time-periodic driving by discretizing space to obtain a Markovian approximation to the continuum dynamics.
That discrete state Markov dynamics was amenable to spectral computations of the current via large-deviation methods~\cite{strand2020current}.
Here, we leverage TDVP with the BTTN architecture to extend that approach to compute the current generated by multiple interacting particles moving in 1D lattice subject to a time-dependent ratcheting potential.
Our companion letter discusses the physics of the ratchet problem in more detail; here we focus on the technical details that allow the TDVP/BTTN methodology to compute the statistical properties of currents in time-periodic steady states~\cite{rahav2008directed,raz2016mimicking,rotskoff2017mapping,barato2018current} in the presence of interactions~\cite{asban2014no}.
In particular, we illustrate that the calculations recapitulate Monte Carlo samples generated by the Gillespie algorithm, while avoiding the sampling noise.

\section{Methods}

\subsection{1D ratchet model}
\label{sec:model}

We set out to study a discretized 1D flashing ratchet with periodic boundary conditions, a tunable driving frequency, and a variable number of particles that interact through volume exclusion.
Many prior investigations of single-particle 1D and 2D ratchets motivated the particular form of the ratcheting potentials~\cite{dasilva2008reversible,mcdermott2016collective,kedem2017drive,kedem2017mechanisms,strand2020current}.
The focus on interactions has some precedent.
Of particular note, Kedem et al.\ have simulated trajectories of many electrons interacting via a Coulomb potential in a 2D ratchet~\cite{kedem2019cooperative}.
As in that work, our transported particles are subject to a spaciotemporal potential
\begin{equation}
    U(x,t)=X(x)T(t),
\end{equation}
where \(T(t)\) and \(X(x)\) are periodic in time and in space, respectively.
For the flashing ratchet model,
\(T(t)\) is a square wave with period \(\tau\) and amplitude \(V_{\text{max}}\) that toggles between \textit{on} and \textit{off} stages:
\begin{equation}
T(t) = \begin{cases}
    -V_{\text{max}}, & 0 \leq t < \frac{\tau}{2}\\
    0, & \frac{\tau}{2} \leq t < \tau.
    \end{cases}
\end{equation}
Following the setup from \cite{strand2020current}, the spatial potential is biharmonic and defined as
\begin{equation}
    X(x) = \frac{a_1}{2}\sin\left(\frac{2\pi x}{x_\text{max}}\right)+\frac{a_2}{2}\sin\left(\frac{4\pi x}{x_\text{max}}\right),
    \label{eq:Xx}
\end{equation}
where \(x_\text{max}\) is the length of the repeating unit and \(a_1\) and \(a_2\) sculpt the potential.
We assign \(a_1 = 1\) and \(a_2 = 0.25\).

\begin{figure}[htb]
    \centering
    \includegraphics[width=0.45\textwidth]{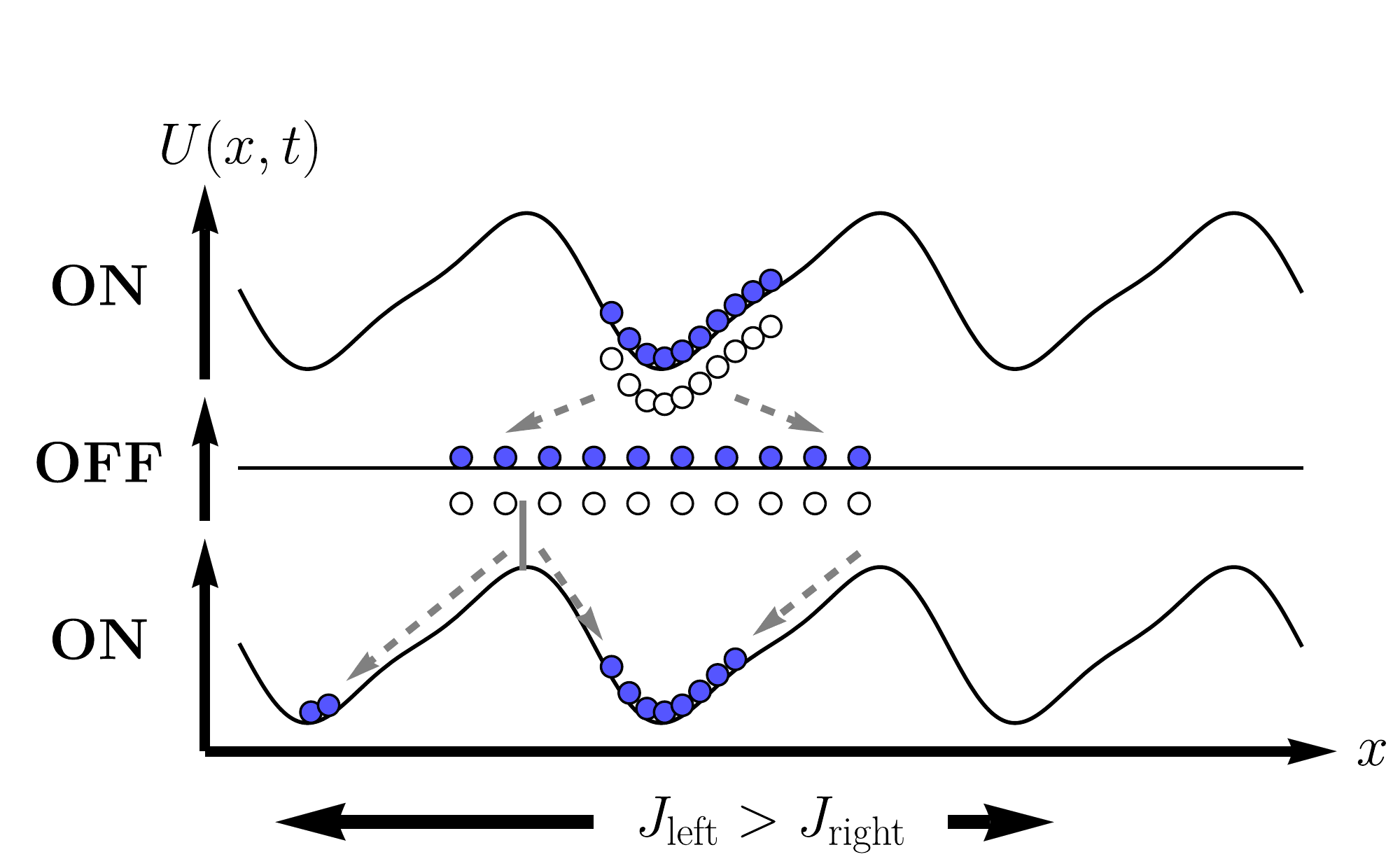}
    \caption{Schematic of the 1D flashing ratchet.
      The driving protocol consists of an potential energy landscape that toggles between the \textit{on} and \textit{off} states with period \(\tau\).
      From the initial time to \(\tau / 2\), particles concentrate in the bottom of a well.
      During the subsequent half period, those particles diffuse outward on a flat landscape before again settling in the wells during the next period.
      The asymmetric potential generates current in the \(-x\) direction, with the current magnitude depending on the frequency of flashing.}
      \label{model}
\end{figure}

The setup is very similar to our prior single-particle work~\cite{strand2020current}, so we highlight two important distinctions.
First, we are now considering a 1D ratchet with particles that can only move along the \(x\) direction; particles in our earlier work generated current along that same \(x\) direction but could additionally move along another dimension.
Second, our temporal function \(T(t)\) toggles between \(V_{\rm max}\) and 0, not between \(V_{\rm max}\) and \(-V_{\rm max}\).
For a 1D ratchet, this move from a symmetric square wave temporal drive to a flashing ratchet is needed to generate nonvanishing current~\cite{reimann2002brownian}.
Otherwise, any motion occurring within the first half of the period would be offset by motion in the opposite direction during the second half of the period.
As illustrated in Fig.~\ref{model}, the flashing ratchet generates current in the negative direction, owing to the asymmetric sawtooth form of \(U(x,t)\).

\subsection{Time-periodic steady-state currents}
\label{sec:currents}

By coarse graining the 1D ratchet in space, the dynamics is modeled as a nearest-neighbor Markov jump process on a periodically replicated lattice of \(N\) sites with grid spacing \(h\).
That jump process obeys the master equation
\begin{equation}
    \frac{\partial|p\rangle}{\partial t}=\mathsf W(t)|p\rangle,
    \label{eq:master}
\end{equation}
where \(\mathsf W\) is the time-dependent rate operator and \(|p\rangle\) is the state vector consisting of the probabilities of each possible system configuration.
The time-dependent \(\mathsf W\) toggles between two distinct sets of rates with a period \(\tau\):
\begin{equation}
    \mathsf{W}(t) = \begin{cases} \mathsf{W}_1, & 0 \leq t < \tau/2,\\ \mathsf{W}_2, & \tau/2 \leq t < \tau. \end{cases}
    \label{eq:ratemat}
\end{equation}
In the first half of the period, dynamics evolves on the sawtooth landscape analogous to \cite{strand2020current}, so \(\mathsf W_1\) is a rate matrix permitting nearest neighbor hops from site \(i\) to site \(i \pm 1\) with rates
\begin{align}
    r_{1,i\rightarrow i\pm1}&= \pm \frac{V_{\rm max} X'(x)}{2h}+\frac{D}{h^2}
\end{align}
provided site \(i \pm 1\) is vacant.
In the continuum \(h \to 0\) limit, the parameter \(D\) becomes the diffusion constant of the associated overdamped single-particle Langevin dynamics~\cite{gingrich2017inferring}.
In the second half of the period, the potential is turned off and the evolution proceeds on a flat landscape.
The rate matrix \(\mathsf W_2\) permits the same volume-excluding nearest neighbor hops, but the rates of those hops are now \(r_{2, i \to i \pm 1} = D / h^2\).
Consistent with the companion paper, we set \(V_\text{max}\) to 0.1~\si{\V} and \(D\) is given the value \(12.64~\si{\um\squared\per\ms}\)~\cite{strand2021companion}.

The long-time limit of Eq.~\eqref{eq:master} approaches the time-periodic steady-state vector \(|\pi\rangle_{t}\) on the time interval \(t \in [0, \tau]\).
When the operators \(\mathsf W_1\) and \(\mathsf W_2\) are cast as matrices, \(|\pi\rangle_0\) is simply obtained as the leading eigenvector of the full-period transition matrix \(\mathsf{T}\equiv e^{\tau\mathsf{W}_2/2}e^{\tau\mathsf{W}_1/2}\).
In this work, we seek period-averaged macroscopic currents around the ring, constructed in terms of the time-dependent currents from site \(\nu\) to site \(\mu\), \(j_{\mu \nu}(t)\):
\begin{equation}
    \bar{\jmath}=\frac{1}{\tau}\int_0^\tau\text{d}t\,\sum_{\mu, \nu}d_{\mu \nu} j_{\mu \nu}(t),
    \label{eq:jd}
\end{equation}
where the weights
\begin{equation}
    d_{\mu \nu}=\begin{cases} +1,\quad &\text{\(\nu\) directly left of \(\mu\),}\\
    -1,\quad&\text{\(\nu\) directly right of \(\mu\),}\\
    0,\quad&\text{otherwise} \end{cases}
\end{equation}
pick out the oriented nearest-neighbor transitions.
To characterize the mean and variance of these currents at the time-periodic steady state, we define the scaled cumulant-generating function (SCGF) \(\psi(\lambda)\) as
\begin{equation}
    \psi(\lambda) := \lim_{n \to \infty} \frac{1}{n} \ln \langle e^{\lambda n \bar{\jmath}}\rangle_n,
\end{equation}
where \(n\) is the number of driving periods.
The first and second derivatives of \(\psi(\lambda)\), evaluated at \(\lambda = 0\), yield the mean and variance of the current~\cite{touchette2009large}.
It is known that \(\psi(\lambda)\) can be obtained from the largest eigenvalue of a product of matrix exponentials~\cite{lebowitz1999gallavotti, lecomte2007thermodynamic, touchette2009large, chabane2020periodically} as
\begin{equation}
    \psi(\lambda) = \frac{1}{\tau} \ln \max {\rm eig} \left(e^{\mathsf{W}_2(\lambda) \tau / 2} e^{\mathsf{W}_1(\lambda) \tau / 2}\right),
    \label{eq:perron}
\end{equation}
with the so-called tilted rate operators \(\mathsf W_k(\lambda)\) defined in terms of the original Eq.~\eqref{eq:ratemat} rate operators as 
\begin{equation}
    [\mathsf W_k(\lambda)]_{\mu \nu}:=[\mathsf W_k]_{\mu \nu}e^{\lambda d_{\mu \nu}}.
\end{equation}

When evolving dynamics of many interacting particles, the matrix representation of the titled operator becomes untenable due to the exponential growth of the state space.
Therefore it is impractical to directly compute the product of matrix exponentials in Eq.~\eqref{eq:perron}.
Instead, we can start with an arbitrary state vector at time zero.
That initial state can be numerically propagated in time by \(\mathsf{W}_1(\lambda)\) for half a period then propagated by \(\mathsf{W}_2(\lambda)\) for another half a period.
This time propagation is continued until the time-periodic steady state is reached, at which point the SCGF is deduced from
\begin{equation}
  |\pi(\lambda)\rangle_{\tau}=\exp(\psi(\lambda)\tau)|\pi(\lambda)\rangle_{0},
    \label{eq:evolution}
\end{equation}
with \(\left| \pi(\lambda)\right>_t\) being the time-periodic steady state subject to exponential bias \(\lambda\).
The advantage of this dynamical approach is that it can be practically implemented for many-body dynamics when the time evolution is approximated by the TDVP algorithm.
That algorithm, which projects the natural dynamics onto a subspace defined by a tensor network (TN) ansatz, leverages the expression of \(\mathsf W_k(\lambda)\) in terms of local operators acting on each lattice site.
In this occupation basis, or \emph{second quantized form},
\begin{align}
\nonumber    \mathsf W_k(\lambda)=&\sum_{i=1}^Nr_{k,i\rightarrow i+1}(e^\lambda \mathbf a_i\mathbf a_{i+1}^\dagger-\mathbf n_i\mathbf v_{i+1})\\
    +&\sum_{i=1}^Nr_{k,i+1\rightarrow i}(e^{-\lambda}\mathbf a_i^\dagger \mathbf a_{i+1}-\mathbf v_i\mathbf n_{i+1}),
    \label{eq:secondquant}
\end{align}
where \(\mathbf a_i\), \(\mathbf a_i^\dagger\), \(\mathbf n_i\), and \(\mathbf v_i\) are annihilation, creation, particle number, and vacancy number operators at site \(i\), respectively.
Note that the periodic boundary conditions lead us to associate \(N+1 \equiv 1\).
Because the second quantized operator only involves nearest-neighbor interactions, it can be expressed as product of operator-valued matrices, one per site of the lattice.
Those operator-valued matrices, discussed explicitly in Appendix~\ref{sec:mpo} and symbolically represented by gray circles in Fig.~\ref{diagram}a, allow Eq.~\eqref{eq:secondquant} to be efficiently computed as the product of the operator-valued matrices, a matrix product operator (MPO)~\cite{schollwock2011density}.
If the state \(\left|p\right>\) is similarly decomposed into a local site representation, the action of \(\mathsf W_k(\lambda)\) on \(\left|p\right>\) can be calculated even when the matrix form of \(\mathsf W_k(\lambda)\) is too large to explicitly construct.
For example, \(\mathsf W_k(\lambda)\) would be a roughly \(10^9 \times 10^9\) matrix for the 32-site lattice with 16 particles discussed in Results.

\subsection{Tensor networks}
\label{sec:structure}
To leverage the MPO, we express the state vector \(|p\rangle\) from Eq.~\eqref{eq:master} in terms of a product basis of local basis states \(|s_i\rangle\) on site \(i\) as
\begin{equation}
    |p\rangle=\sum_{s_1\cdots s_N}c_{s_1\cdots s_N}|s_1\cdots s_N\rangle.
    \label{eq:productstate}
\end{equation}
The rank-\(N\) tensor \(c\) depends on the \(N\) physical indices \(s_1, \cdots s_N\), but \(c\) is so high dimensional that it cannot be practically computed.
  Rather, we introduce a tensor network (TN) ansatz in which \(c\) is generated by a network of tensors \(A^{(1)}, A^{(2)}, \hdots A^{(\chi)}\), each with modest rank.
  Those tensors can depend on some of the physical indices (\(s_1, \hdots, s_N\)) reflecting the occupation at each site as well as some auxiliary indices that will be summed over.
  We adopt the nomenclature that \(S_i\) is a set of physical indices upon which the \(i^{\rm th}\) tensor depends (potentially an empty set) and \(Q_i\) the set of auxiliary indices.
  The tensor product ansatz is a restriction that we only allow expansion coefficients of the form
  \begin{equation}
    c_{s_1\hdots s_N} = \sum_{Q_1 \hdots Q_{\chi}} A_{Q_1, S_1}^{(1)} A_{Q_2,S_2}^{(2)} \hdots A_{Q_{\chi}, S_{\chi}}^{(\chi)},
    \label{eq:ansatz}
  \end{equation}
  yielding a state vector that is parameterized by the set of all \(A\)'s as
  \begin{equation}
    |p[A]\rangle=\sum_{\substack{S_1\cdots S_\chi\\Q_1\cdots Q_\chi}}A_{Q_1, S_1}^{(1)} A_{Q_2, S_2}^{(2)}\cdots A_{Q_\chi, S_{\chi}}^{(\chi)}|s_1\cdots s_N\rangle.
    \label{eq:gentn}
  \end{equation}
  Note that any choice of tensors \([A]\) will yield a rank-\(N\) tensor \(c\), but given an arbitrary \(c\) it might not be possible to express exactly it exactly in terms of a set \([A]\) 
  Indeed, the realization of any arbitrary rank-\(N\) tensor in terms of a TN requires that the auxiliary indices linking the tensors are sufficiently high dimensional.
  The TN ansatz derives its computational utility by restricting that auxiliary index dimensionality, the so-called bond dimension.
  By finding a bond dimension which is large enough but not too large, it is often possible to make a good approximation to the exact dynamics while gaining the computational benefit of low dimensional tensors.
  Specifically, we cap the bond dimension at \(m\), a tunable variational parameter, that generally controls how much the auxiliary indices can mediate correlations between nearby physical indices.
  Too large an \(m\) inevitably renders TN calculations intractable, whereas too small of an \(m\) generates an inflexible subspace on which variational calculations are excessively constrained.
  Capping the bond dimension necessarily means that one discards some information, so as we will discuss, singular value decompositions (SVD) are strategically employed to preserve only the \(m\) most essential components of a matrix.

  \section{Dynamics of the tensor network}

  \subsection{The Binary Tree Tensor Network}
  \label{sec:TheBTTN}
  To actually perform a calculation, it is necessary to specialize to a particular design of how tensors are connected in a network; in other words, one must specify which indices belong to each of the sets \(S_i\) and \(Q_i\).
  For 1D quantum and classical systems, the choice of network is usually a MPS.
  That MPS ansatz has proved to be convenient and robust for many applications~\cite{schollwock2011density}. The convenience derives from the ability to generate a canonical form or Schmidt decomposition, which allows for efficient and stable computations on an MPS~\cite{vidal2003efficient}.
  Unfortunately, for systems with periodic boundary conditions, it is not possible to represent an MPS in a canonical form due to the loop in the TN structure~\cite{schollwock2011density}.
  To handle the ratchet's periodic boundary conditions with a loopless TN that supports a canonical form, we therefore use a BTTN~\cite{Murg2010,Shi2006}.

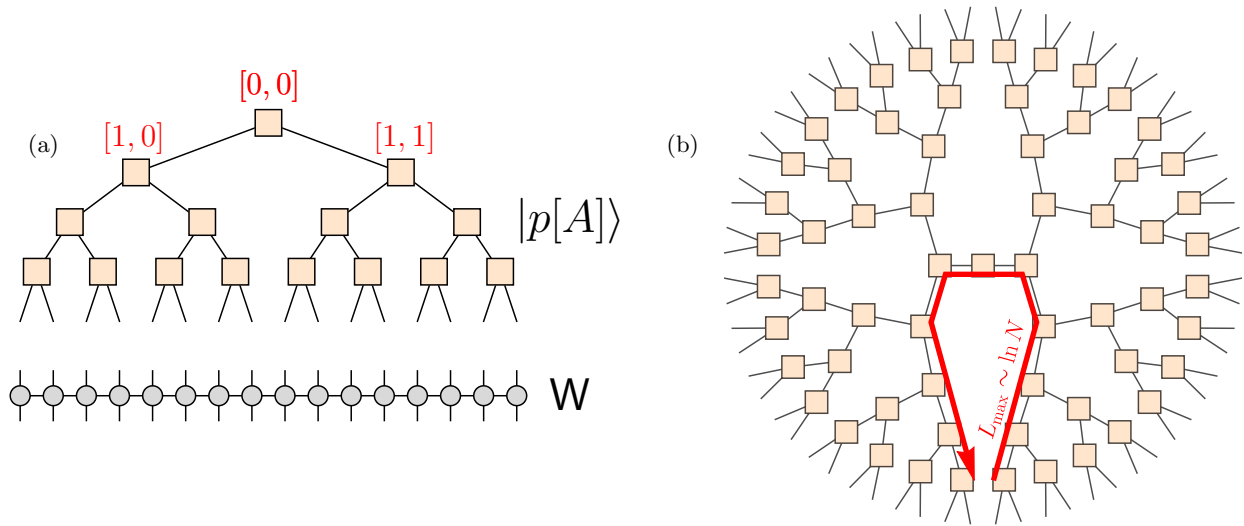
\begin{figure*}[htb]
    \centering
    \begin{tikzpicture}
        \node at (-1,6.4) {\includegraphics[width=0.4\textwidth]{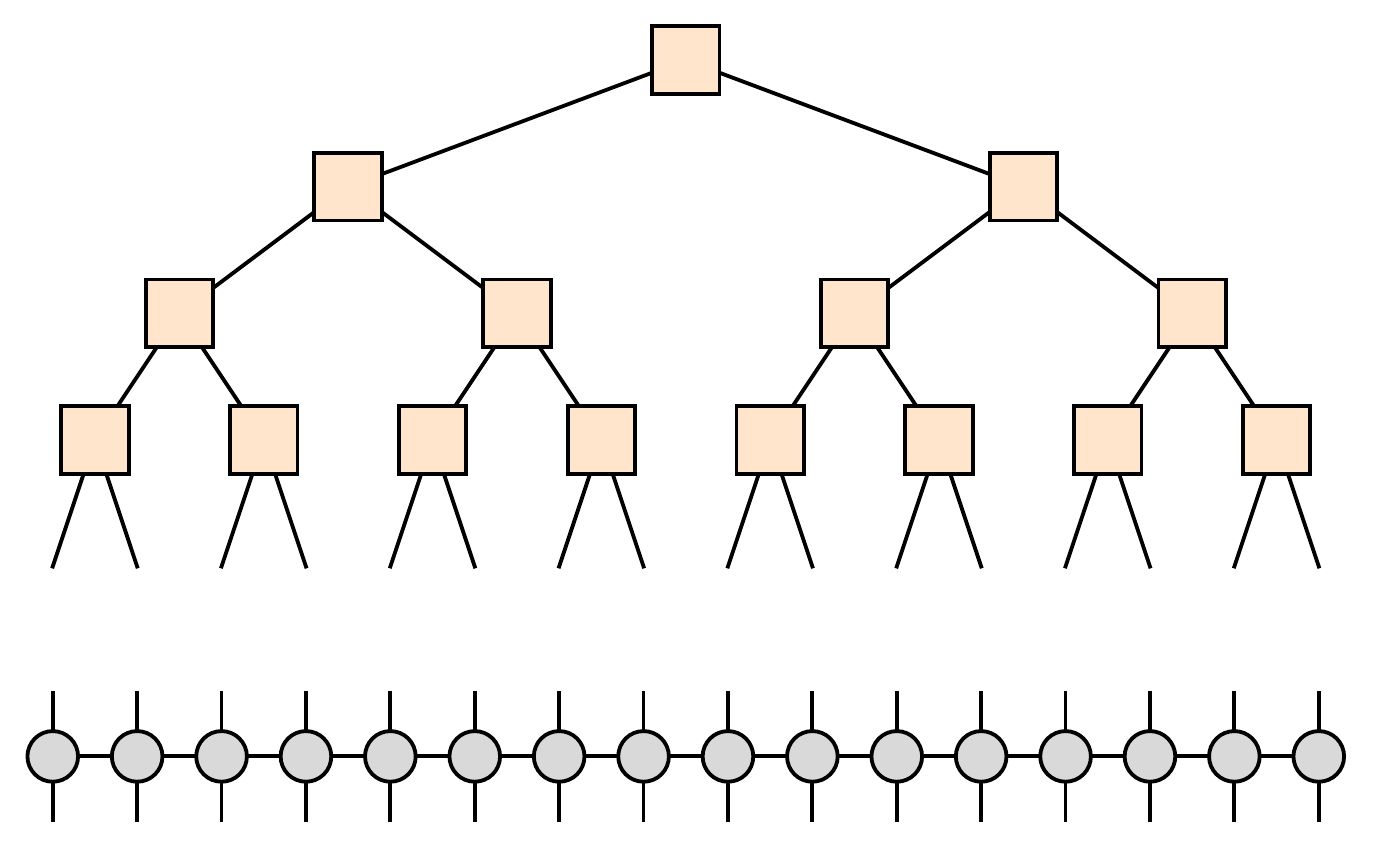}};
        \node at (8.5,6.4) {\includegraphics[width=0.4\textwidth]{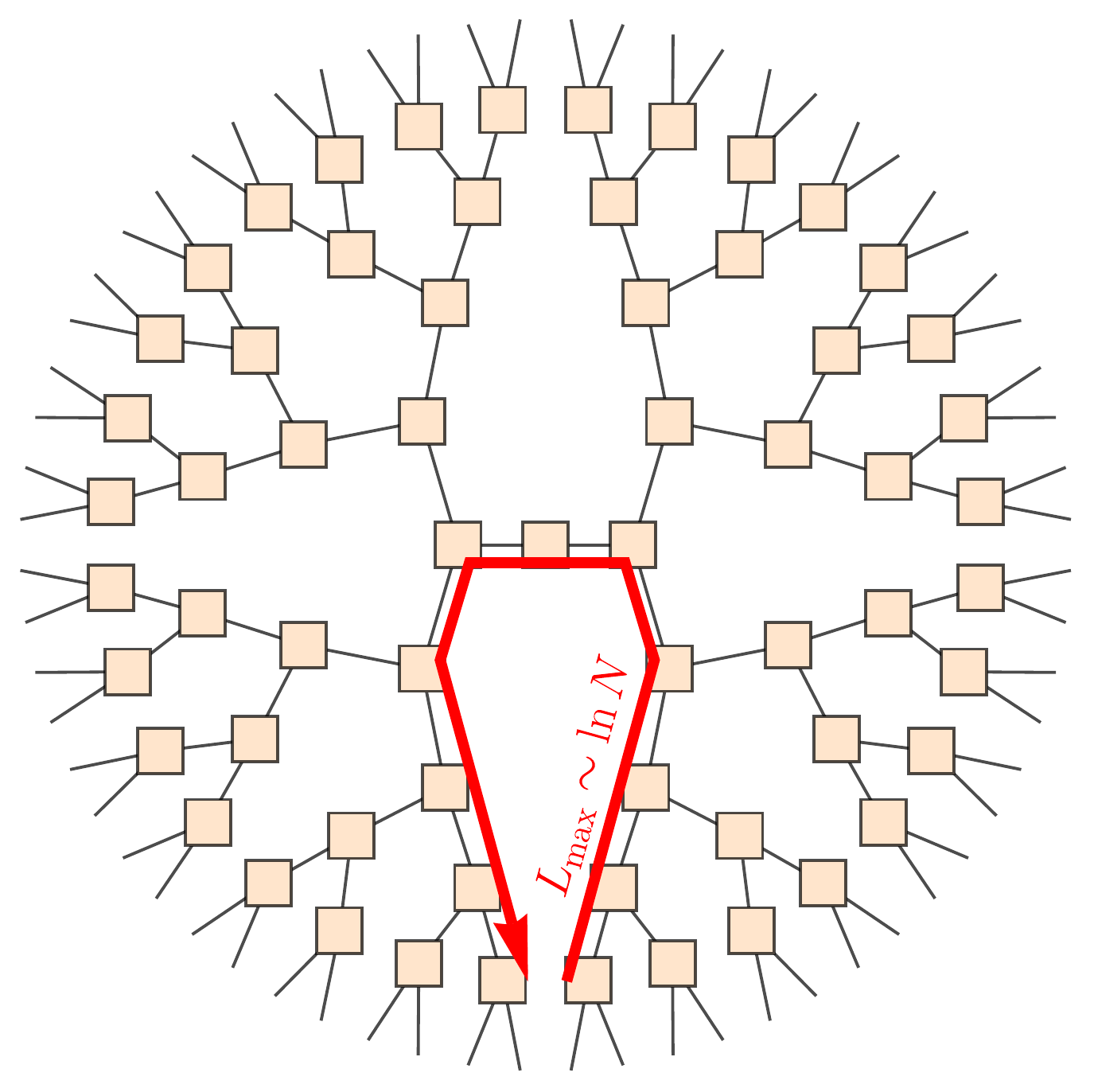}};
        \node at (-4,8) {(a)};
        \node at (4.5,8) {(b)};
        \node at (3,7) {\LARGE\(|p[A]\rangle\)};
        \node at (3,4.7) {\LARGE\(\mathsf W\)};
        \node at (-1,8.8) {\color{red}\large\([0,0]\)};
        \node at (-1,8.8) {\color{red}\large\([0,0]\)};
        \node at (-2.8,8.1) {\color{red}\large\([1,0]\)};
        \node at (0.8,8.1) {\color{red}\large\([1,1]\)};
    \end{tikzpicture}
    \caption{Binary tree tensor network (BTTN) diagrams.
(a) This BTTN corresponds to a 16-site lattice, with tensors and indices shown as squares and lines, respectively.
The lines sticking out of the bottom layer of the tree represent physical indices and can contract with the corresponding physical indices in a matrix product operator (MPO), representing some rate operator \(\mathsf W\).
This diagram therefore illustrates the action of \(\mathsf W\) on an arbitrary state \(|p\rangle\).
(b) Here, the BTTN is built from a 64-site lattice and has tensors visually rearranged to emphasize the one-to-one mapping of tensors to sites along a circular lattice, as found in the 1D ratchet studied in this work and any other system subject to periodic boundary conditions.
A red arrow is added to depict the largest distance \(L_\text{max}\) one has to travel between any two neighboring lattice sites.
Contrary to a loopless MPS, here \(L_\text{max}\) scales logarithmically with the number of sites, thus rendering BTTN methods both accurate and tractable even with the absence of loops.}
    \label{diagram}
\end{figure*}

  The tree itself is illustrated in Fig.~\ref{diagram}a.
Following~\cite{gerster2014unconstrained}, we label each tensor \(A^{(i)}\) not by a single superscript \((i)\) as in Eq.~\eqref{eq:ansatz}, but rather by the pair \([l, i]\) indicating that the tensor appears in \(i^{\rm th}\) node of the \(l^{\rm th}\) layer of the tree.
These \(L \equiv \log_2 N\) layers count up from 0 at the root of the tree to \(L - 1\) at the base while the sites count up from 0 to \(2^l-1\) moving from left to right across a layer.
Into the base of the tree feed \(N\) physical indices with dimension \(d\) = 2 corresponding to lattice sites which are either occupied or unoccupied.
Those tensors of the \(l = L-1\) layer feed upward into parent tensors via auxiliary links.
To capture all possible rank-\(N\) tensors \(c\), the dimension of each auxiliary index must grow such that the link between layers \(l\) and \(l+1\) would have dimension \(M(l) = 2^{2^{L-l-1}}\).
Assuming auxiliary indices are truncated at a maximum bond dimension \(m\), the auxiliary link between \(l\) and \(l+1\) actually has dimension \(\min(m, M(l))\).

The tree structure offers two principle benefits.
Its loopless structure provides access to a canonical form, dramatically simplifying calculations.
Furthermore, the BTTN allows correlations between pairs of lattice sites since each physical index is connected to each other physical index by a pathway whose length grows only logarithmically with the number of lattice sites (see Fig.~\ref{diagram}b).
Due to these merits, the BTTN has been applied to compute ground states via DMRG~\cite{gerster2014unconstrained} and dynamics via TDVP~\cite{kohn2020superfluid,bauernfeind2020time}.
We follow those works closely in applying the methodology to our problem.
  
\subsection{Orthogonalization of the BTTN}
\label{sec:ortho}
\begin{figure*}[htb]
    \centering
    \begin{tikzpicture}
        \node at (0,7.8) {\includegraphics[width=0.23\textwidth]{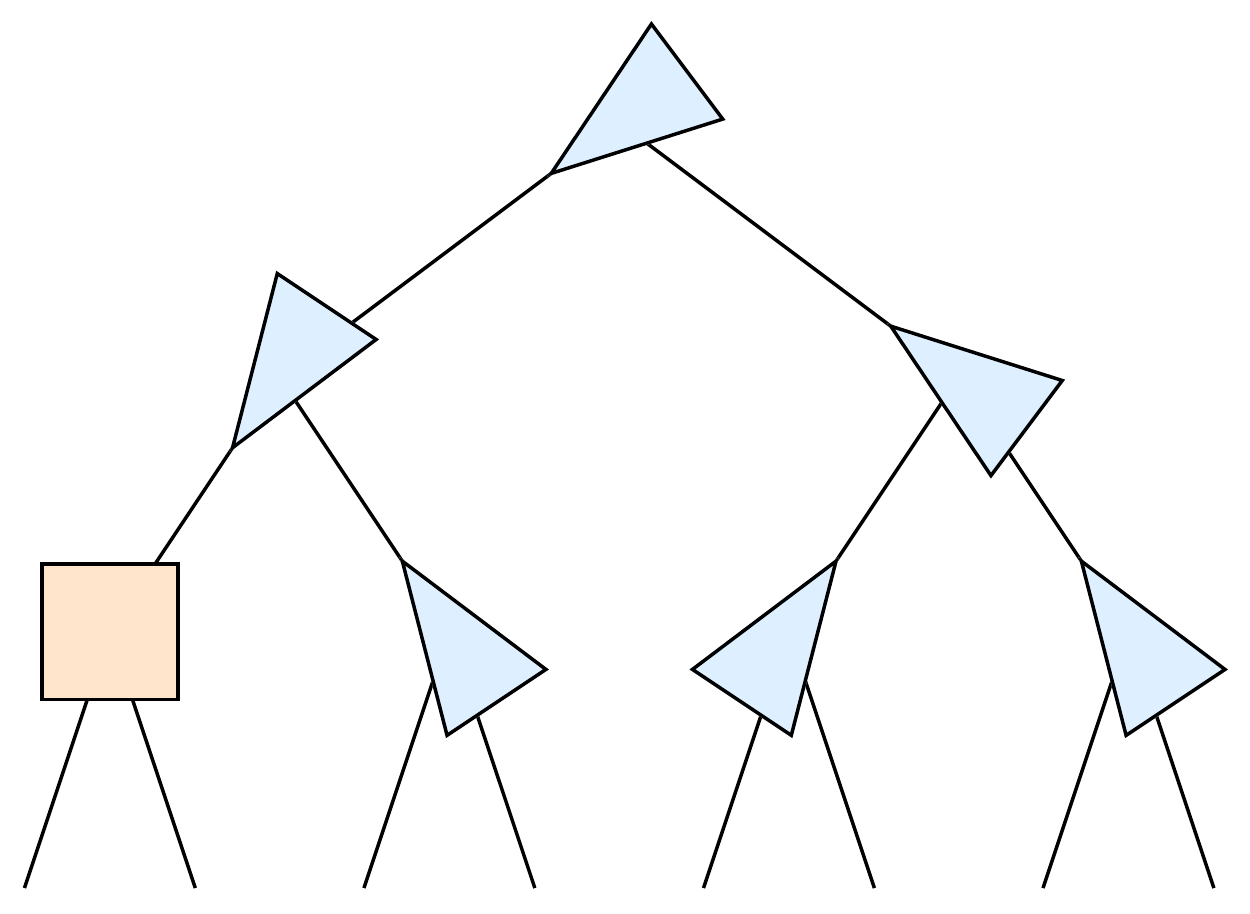}};
        \node at (4.5,7.8) {\includegraphics[width=0.23\textwidth]{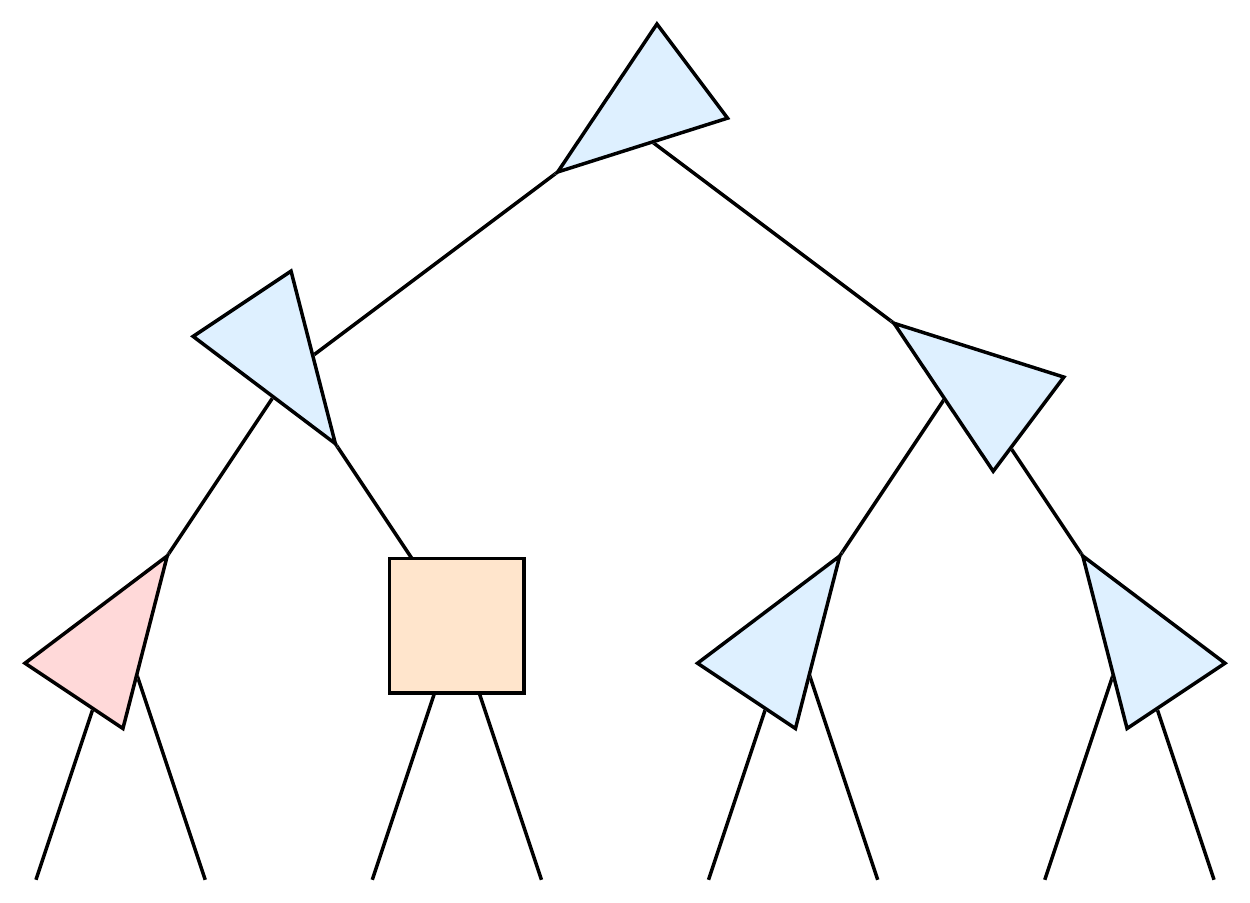}};
        \node at (9,7.8) {\includegraphics[width=0.23\textwidth]{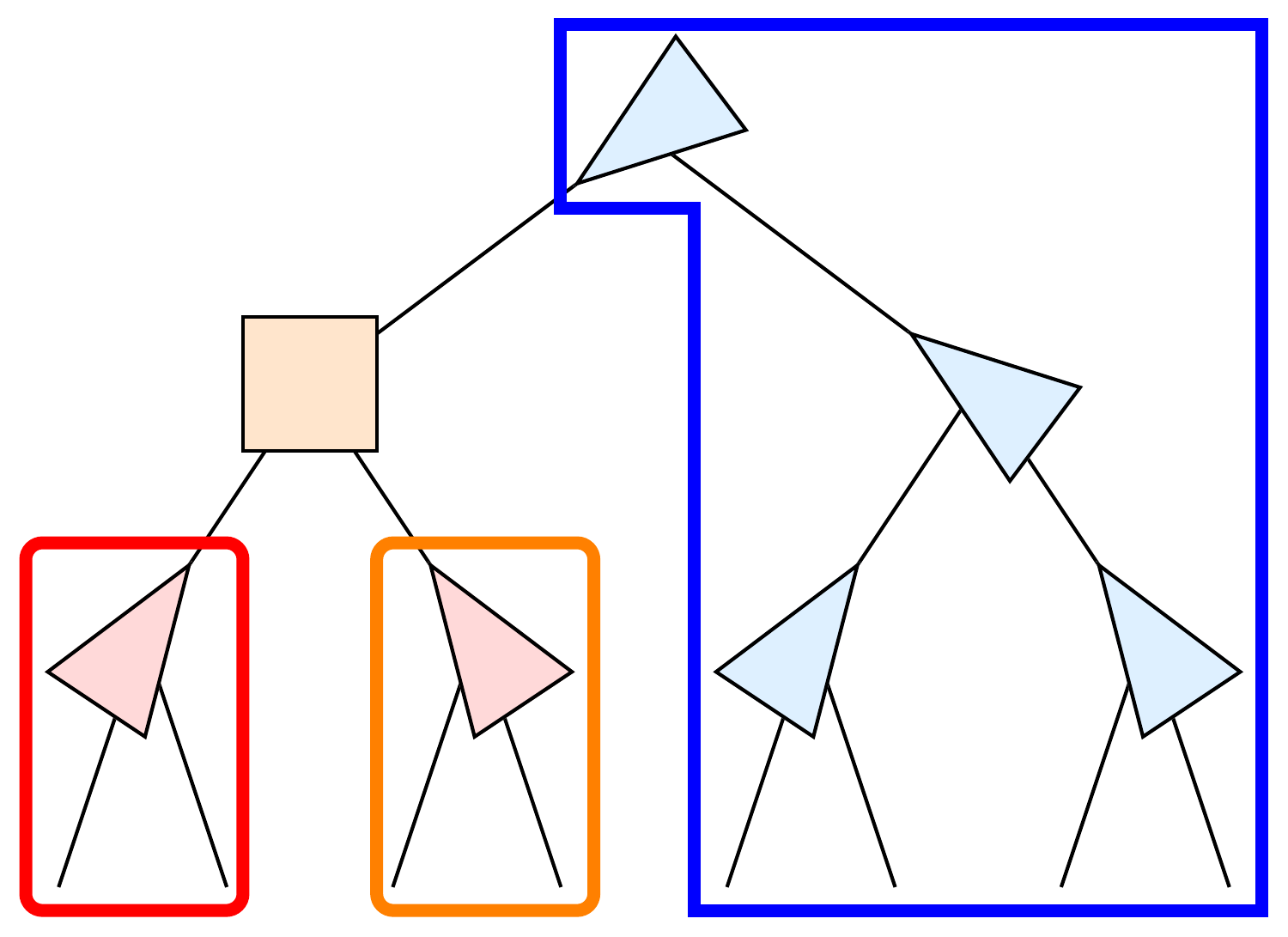}};
        \node at (13.5,7.8) {\includegraphics[width=0.23\textwidth]{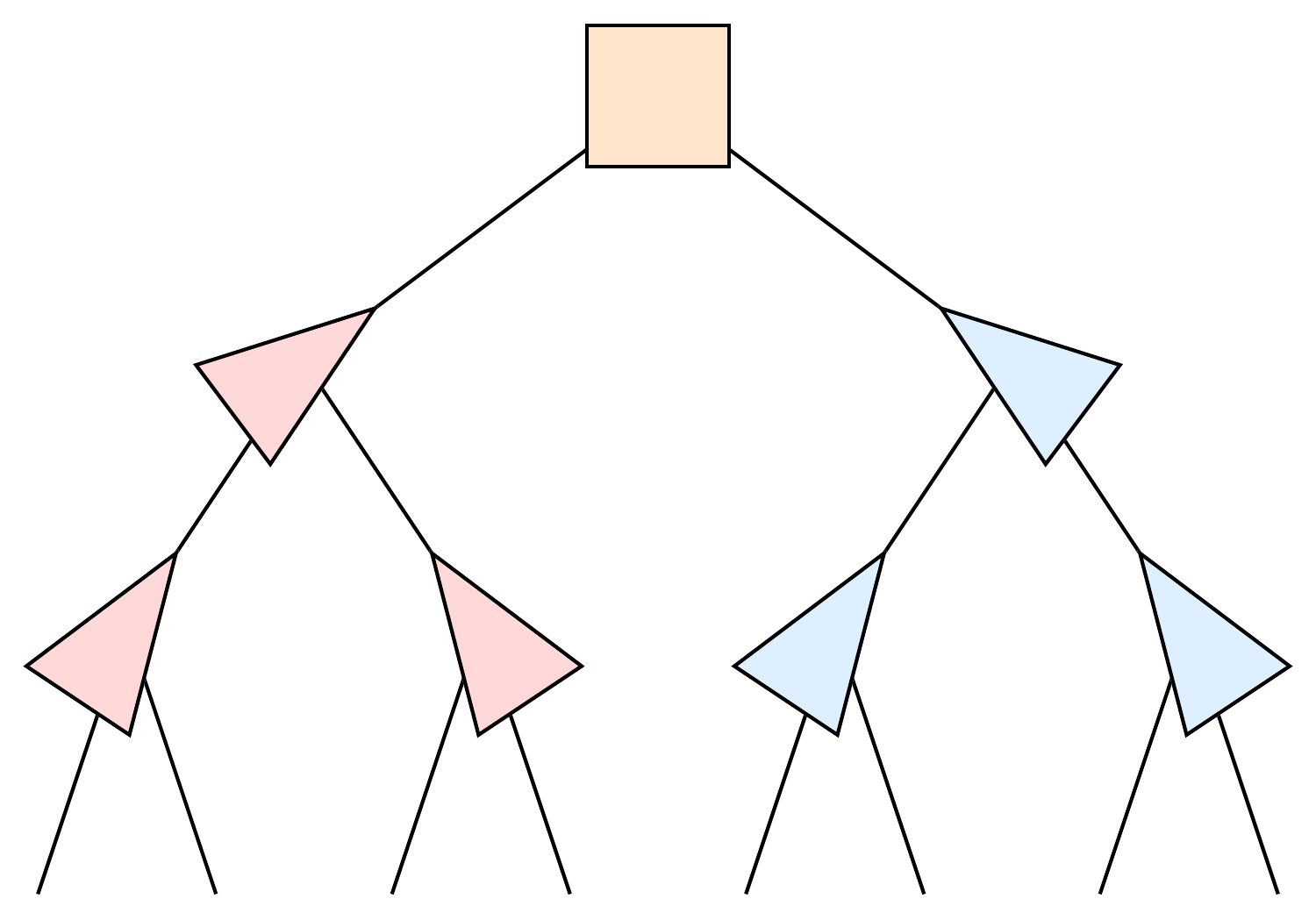}};
        \node at (2.25,4.5) {\includegraphics[width=0.23\textwidth]{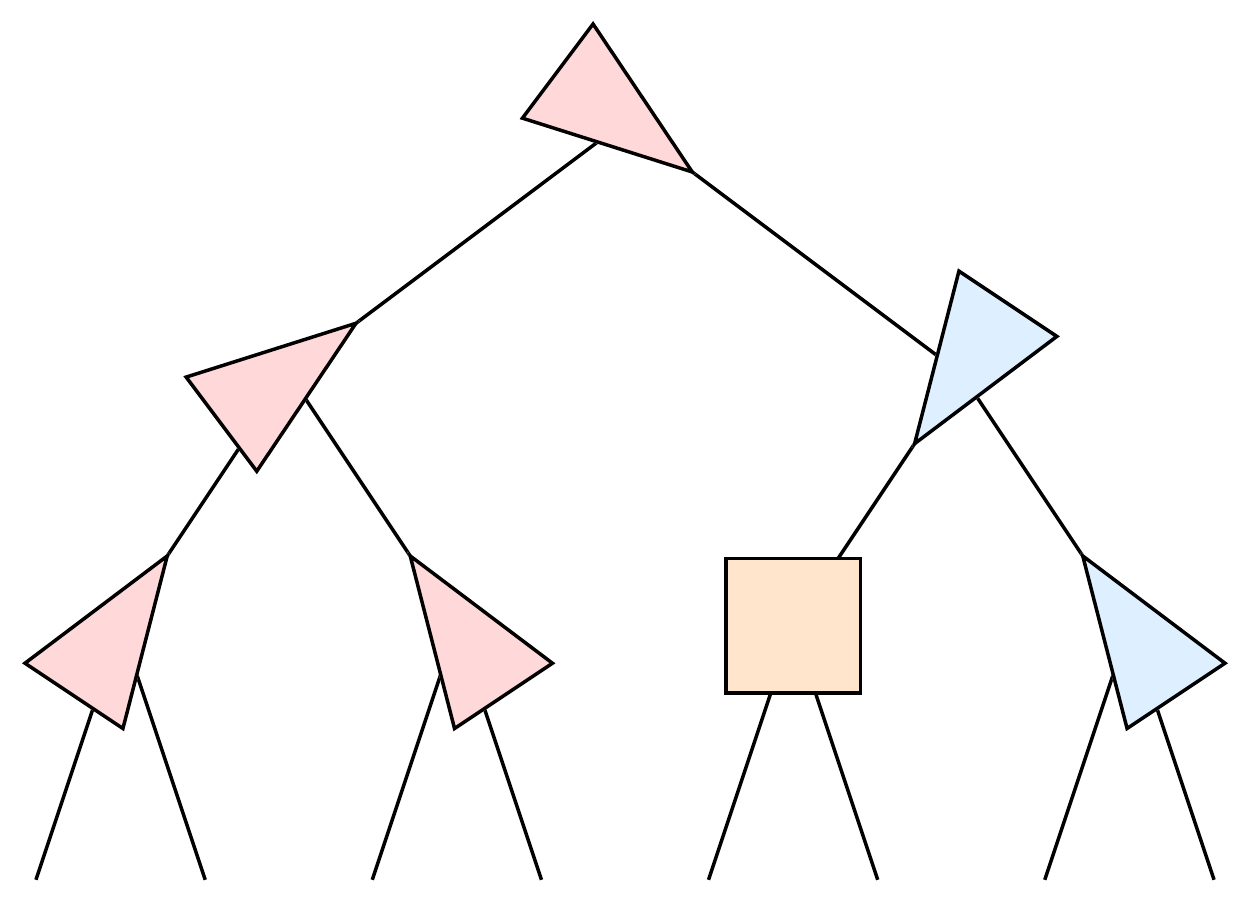}};
        \node at (6.75,4.5) {\includegraphics[width=0.23\textwidth]{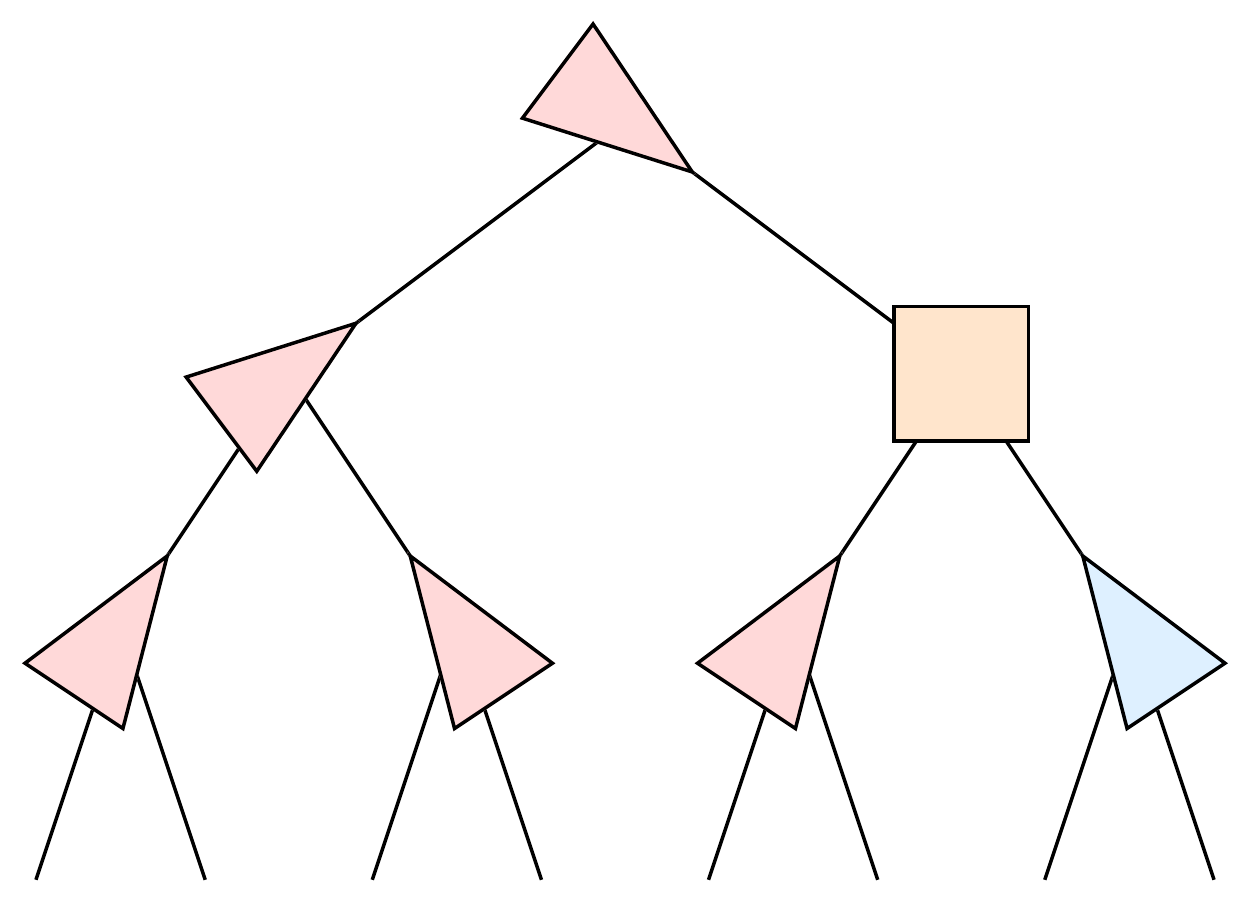}};
        \node at (11.25,4.5) {\includegraphics[width=0.23\textwidth]{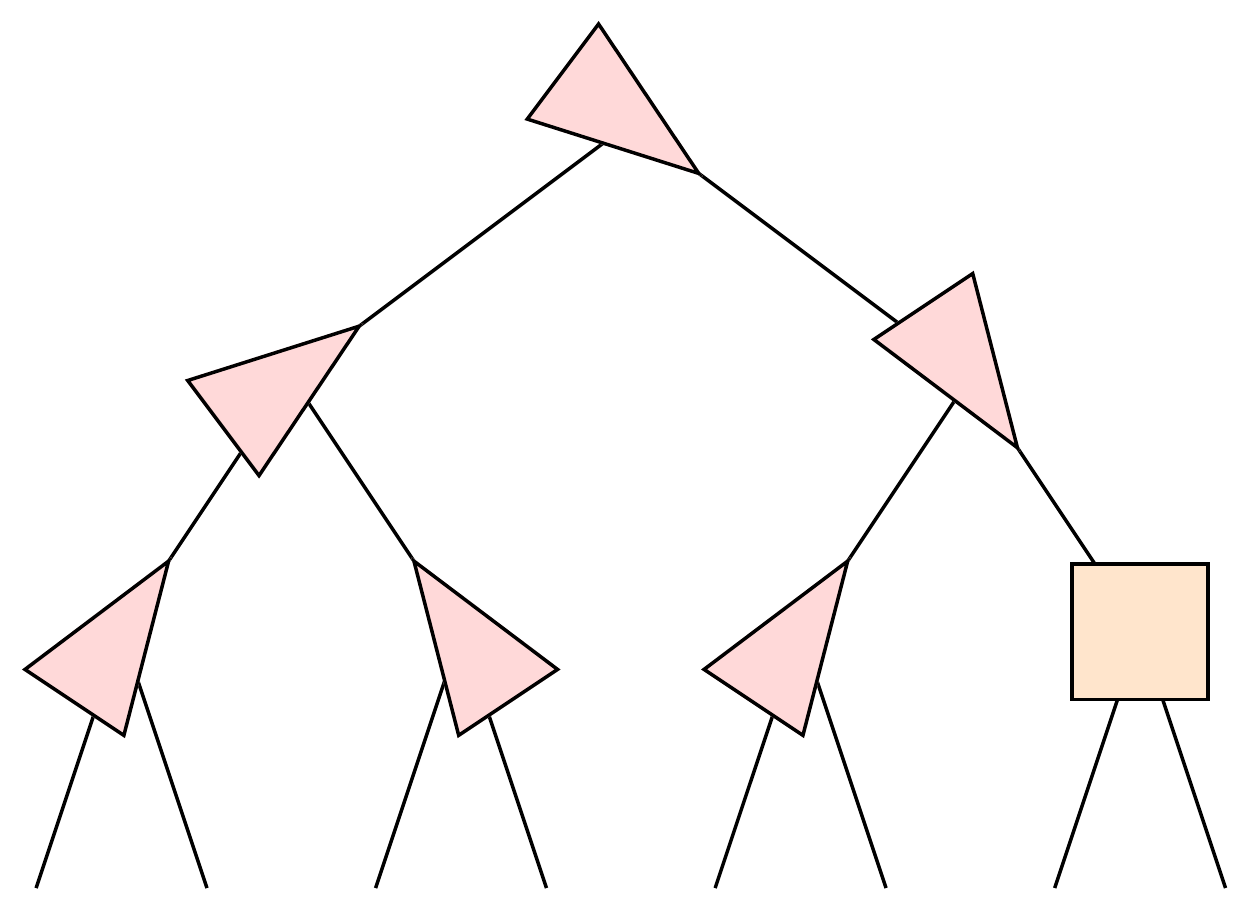}};
        \node at (-1.5,8.6) {(a)};
        \node at (3,8.6) {(b)};
        \node at (7.5,8.6) {(c)};
        \node at (12,8.6) {(d)};
        \node at (0.75,5.3) {(e)};
        \node at (5.25,5.3) {(f)};
        \node at (9.75,5.3) {(g)};
        \node at (8.65,8.1) {\(A^{[1,0]}_{\alpha\beta\gamma}\)};
        \node at (8.4,8.6) {\(\alpha\)};
        \node at (7.5,7.75) {\(\beta\)};
        \node at (8.03, 7.6) {\(\gamma\)};
        \node at (7.4,6.1) {\textcolor{red}{\(\left|L_{\beta}^{[2,0]}\right>\)}};
        \node at (8.5,6.1) {\textcolor{orange}{\(\left|R_{\gamma}^{[2,1]}\right>\)}};
        \node at (10.0,6.1) {\textcolor{blue}{\(\left|P_{\alpha}^{[0,0]}\right>\)}};
    \end{tikzpicture}
    \caption{
    TDVP traversal order ((a) to (g)) for an 8-site lattice's BTTN.
    The starting and ending points are the left-most and right-most leaves of the tree, respectively.
    In each diagram, the tensor currently being updated, namely node \([l,i]\), is colored in beige.
    Tensors which have already been time propagated are colored red and the triangle shapes are used to point at the tensor serving as the orthogonalization center, \([l,i]\).
    Tensors which remain to be time propagated are colored in blue.
    Environment tensors \(\left|P\right>, \left|L\right>,\) and \(\left|R\right>\) are the composition of all tensors in the beige tensor's parent branch, left branch, and right branch, respectively.
    If a sweep is divided into two half-sweeps, as is often seen for MPS methods, initially steps (a) to (g) are performed, followed by a second set of updates in the reverse order.}
    \label{traversal}
\end{figure*}
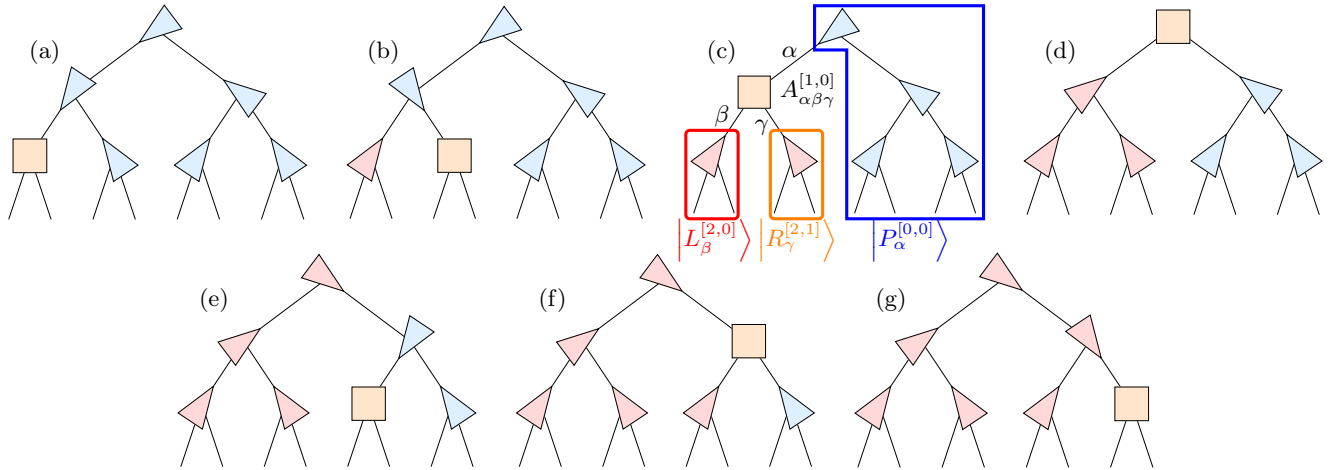

The mapping from the \([A]\) tensors to the expansion coefficient \(c\) is many-to-one, so different combinations of values for the tensors can yield an identical state \(\left|p\right>\).
One way this so-called \textit{gauge freedom} can come about is by introducing a resolution of the identity, \(D^{-1} D\) between tensors at neighboring sites~\cite{Silvi2019}.
If one tensor is transformed by \(D^{-1}\) while its neighbor has a compensatory transformation by \(D\), then the contraction of the tensors is unaffected though the individual tensors will change.
Typically, one leverages the gauge freedom even more aggressively, transforming many tensors in a way that strategically privileges one node \([l,i]\).
Observe that the tensor \(\mathbf A^{[l,i]}\) is linked to one parental branch and two child branches.
It is convenient to contract together all the tensors along a branch to get three so-called environment tensors \(\left|P^{[l-1,i/2]}\right>, \left|L^{[l+1, 2i]}\right>\), and \(\left|R^{[l+1, 2i+1]}\right>\) (see Fig.~\ref{traversal}c), which capture the cumulative effect of the parent branch, left child branch, and right child branch, respectively.
Note that each of these environment tensors depends on a single auxiliary index (one that feeds into \(\mathbf A^{[l,i]}\)) as well as all the physical indices associated with its branch of the tree.
A state \(\left|p\right>\) is orthogonalized about \([l,i]\) when it can be written as
\begin{equation}
    \left|p^{[l,i]}[A]\right>= \sum_{\alpha, \beta, \gamma} A_{\alpha\beta\gamma}^{[l,i]}\left|P_{\alpha}^{[l-1,i/2]}\right>\left|L_{\beta}^{[l+1,2i]}\right>\left|R_{\gamma}^{[l+1,2i+1]}\right>,
    \label{eq:canonicaltree}
\end{equation}
with a gauge chosen such that the environment tensors satisfy the orthonormality conditions \(\langle P_{\alpha'} | P_\alpha\rangle = \delta_{\alpha, \alpha'}\), \(\langle L_{\beta'} | L_\beta\rangle = \delta_{\beta, \beta'}\), and \(\langle R_{\gamma'}| R_\gamma \rangle = \delta_{\gamma, \gamma'}\).
The computational benefit of this chosen gauge is clearest by computing the norm of the BTTN state:
\begin{equation}
    \left\langle p^{[l,i]}[A]\Big|p^{[l,i]}[A]\right\rangle= \sum_{\alpha, \beta, \gamma} A_{\alpha\beta\gamma}^{[l,i]\dagger}A_{\alpha\beta\gamma}^{[l,i]},
\end{equation}
with \(\dagger\) denoting the Hermitian conjugate.
Due to the environment tensor orthonormality, the norm only depends on the tensor at \([l,i]\).

The BTTN TDVP algorithm must advance \(\left|p\right>\) in time by advancing each of the \([A]\) in time, one by one.
Akin to the norm calculation, time evolution of \(\mathbf A^{[l,i]}\) is most efficient if the BTTN has first been orthogonalized about \([l,i]\).
After that propagation of \(\mathbf A^{[l,i]}\), a new gauge transformation can re-orthogonalize about a new node \([l', i']\) to allow  the tensor at that node to be efficiently propagated.
An explicit algorithm to carry out those BTTN orthogonalizations performs successive SVD on all tensors except the orthogonalization center~\cite{Silvi2019,gerster2014unconstrained,bauernfeind2020time}.
For each SVD, a truncation step can be added to respect the maximum bond requirement, the singular values are then sorted and all but the $m$ largest ones are discarded~\cite{schollwock2011density}.

\subsection{Time evolution of tensor network states}
\label{sec:tdvp}

In Section~\ref{sec:currents} we cast the calculation of currents in terms of a dynamics problem, requiring that we propagate a state \(\left|p\right>\) in time with propagators \(\mathsf W_k(\lambda)\).
If we were to represent \(\left|p \right>\) with the full rank-\(N\) tensor as in Eq.~\eqref{eq:productstate}, this time evolution requires that we numerically solve for the time-dependence of the expansion coefficient \(\mathbf{c}\).
In Section~\ref{sec:TheBTTN} we argued that \(\left|p\right>\) should instead be constructed from a set of tensors \([A]\) with a restricted bond dimension.
Imagine propagating this state for time \(\Delta t\) with the tilted operator: \(e^{\mathsf W_k(\lambda) \Delta t} \left|p[A]\right>\).
That newly evolved state generally cannot be exactly constructed in terms of the BTTN with the restricted bond dimension.
Rather, the dynamics that starts with a BTTN state will have left the manifold of BTTN states and leaked into a nearby state in the space of possible rank-\(N\) \(\mathbf{c}\).
The earliest attempts to propagate TN states approximated the matrix exponential with a discrete timestep, but these approaches like the time-evolving block decimation (TEBD)~\cite{vidal2004efficient, verstraete2004matrix} could run into problems associated with the departure from the manifold of TN states~\cite{vidal2007classical,haegeman2011time}.
An alternative approach, first proposed by Dirac and Frenkel as a broad technique for variationally optimized dynamics~\cite{dirac1930exchange, frenkel1934wave}, seeks to propagate \(\left|p \right>\) with the constraint that the state remains confined on a variational manifold of allowed states.
Conceptually, one can think of that constrained dynamics as consisting of the ordinary matrix exponential \(e^{\mathsf W_k(\lambda) \Delta t}\) for a small time \(\Delta t\) immediately followed by a projection onto the variational states.
This TDVP was resurrected by Haegeman et al.\ when they demonstrated that the TDVP approach proved particularly effective when combined with the flexibility of a TN ansatz~\cite{haegeman2011time,haegeman2016unifying}

That TN implementation of TDVP, initially implemented for an MPS but later updated for tree tensor networks~\cite{kohn2020superfluid,bauernfeind2020time}, provides an algorithm to evolve \(\left| p(t)\right>\) with a discrete timestep by computing an equation of motion for the tensors \([A]\) that parametrize the variational state.
The TN ansatz combines especially nicely with the TDVP approach because for a suitably orthogonalized BTTN, the algorithm implementing \([A]\)'s time evolution can efficiently act on one single \(\mathbf A^{[l,i]}\) tensor at a time.
We carried out Bauernfeind et al.'s single-center TDVP procedure~\cite{bauernfeind2020time}, which we describe here.
To avoid truncation errors, we calculated dynamics using BTTN states with a fixed bond dimension, motivating the choice of a single-center algorithm over two-center alternative~\cite{kohn2020superfluid}.

The algorithm starts with a set of tensors \([A]\) at time zero and carries out a step with timestep \(\Delta t\) to yield a new set \([A']\) for that later time.
Tensors in the BTTN are updated one by one according to an ordering for the tree traversal illustrated in Fig.~\ref{traversal}. At tensor \(\mathbf A^{[l,i]}\), the BTTN is first orthogonalized about node \([l,i]\). An effective operator \(\mathsf W_\text{eff}^{[l,i]}\) is then constructed by contracting the MPO with the environment tensors \(\left|P^{[l-1,i/2]}\right>, \left|L^{[l+1, 2i]}\right>\), and \(\left|R^{[l+1, 2i+1]}\right>\), and their conjugate transposes. 
This allows propagation of \(\mathbf A^{[l,i]}\) forward in time for a timestep of \(\Delta t/2\) via a Lanczos exponentiation routine~\cite{hochbruck1997krylov}. 
That node is now said to be evolved forward by \(\Delta t / 2\). The tensor which had just been evolved in time is then decomposed via an SVD into a product of orthogonal unitary operators \(\mathbf{U}\) and \(\mathbf{V}^{\dagger}\) sandwiching a diagonal matrix of singular values \(\mathbf{S}\).
That \(\mathbf{U}\) is retained as the new time-propagated tensor at node \([l,i]\) but the product \(\mathbf{S} \mathbf{V}^{\dagger}\) will be contracted with the neighboring node to shift the orthogonalization center in preparation for the next node of the tree traversal sequence.
Notice that \(\mathbf{S}\mathbf{V}^{\dagger}\), which will be contracted into the neighboring node, was already advanced in time by an extra \(\Delta t / 2\) relative to that neighboring node.
Before contracting them together, it is therefore necessary to propagate \(\mathbf{S} \mathbf{V}^{\dagger}\) backward in time by \(\Delta t / 2\).
The net result is that node \([l,i]\) is advanced by \(\Delta t / 2\) and the orthogonalization center is shifted to the next node in sequence.
That time propagation of a single tensor corresponds, for example, to the jump from Fig.~\ref{traversal}c to Fig.~\ref{traversal}d.
A full timestep is achieved by sweeping through the entire tree to sequentially update the tensors in the order of Fig.~\ref{traversal}.
One pass through the tree propagates the state by \(\Delta t / 2\), then the full timestep \(\Delta t\) is completed by sweeping back through the tree in reverse order.

\section{Results}
\label{sec:results}

\subsection{Constructing the initial BTTN state}
\label{sec:inputseed}

\begin{figure*}[htb]
    \centering
    \begin{tikzpicture}
        \node at (-1,6.4) {\includegraphics[width=0.45\textwidth]{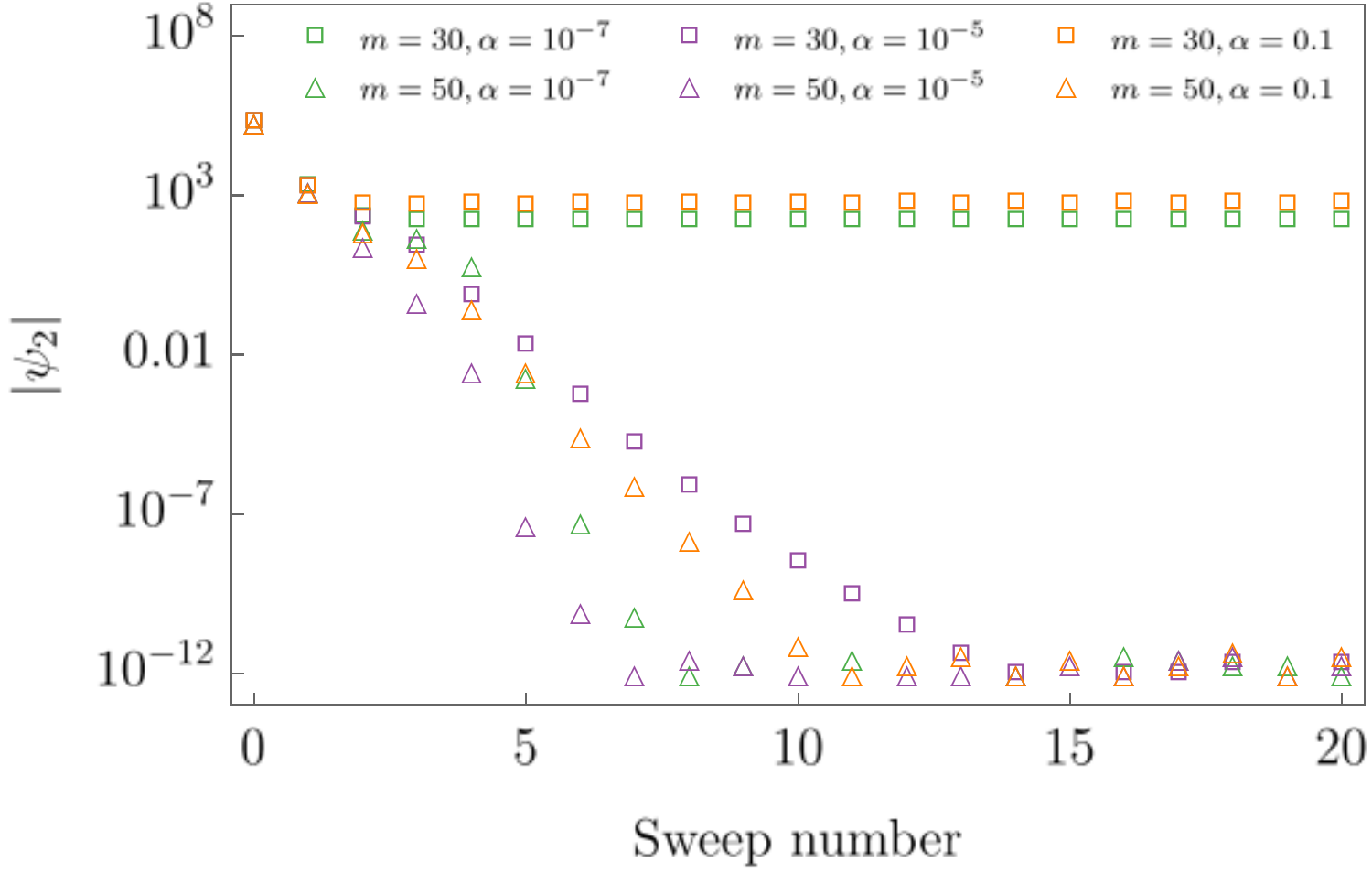}};
        \node at (8.5,6.4) {\includegraphics[width=0.45\textwidth]{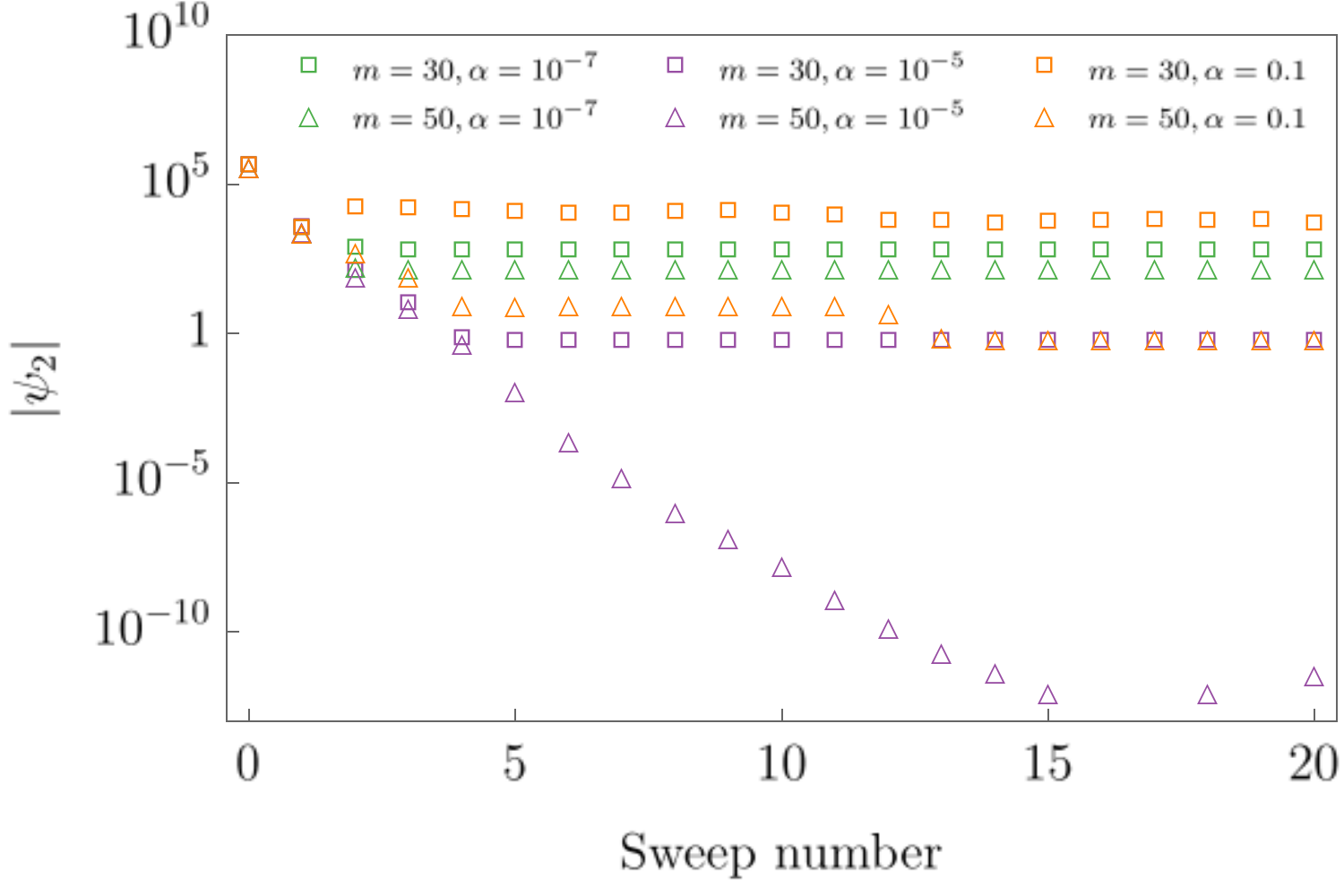}};
    \end{tikzpicture}
    \caption{
    Convergence to the steady state of \(\mathsf{W_2}\), \(\left|\pi_2\right>\), by DMRG to serve as a seed in the TDVP calculations.
    As a rate matrix, the top eigenvalue of \(\mathsf{W_2}\) is zero, so convergence was assessed by monitoring how the estimate of that top eigenvalue \(\psi_2\) approached zero for a 32-site lattice with 8 (left) and 16 (right) particles.
    The DMRG calculations were repeated with maximal bond dimension \(m\) of 30 (squares) and 50 (triangles) and with subspace expansion mixing parameter \(\alpha\) of \(10^{-7}\), \(10^{-5}\), and \(0.1\) (green, purple, and orange, respectively).
    For the DMRG to fully converge, \(m\) must be sufficiently large and \(\alpha\) must be neither too large nor too small.}
    \label{dmrgconv}
\end{figure*}

To compute the SCGF for currents using the tilted dynamics of Eq.~\eqref{eq:evolution}, we first must generate an initial BTTN state.
That initial state should satisfy two needs.
Firstly, it should be similar to the time-periodic steady state.
By the Perron--Frobenius theorem, an arbitrary initial state would relax into the time-periodic steady state, but the closer the initial state is, the faster TDVP can converge.
Secondly, the initial state must be constructed with a maximal bond dimension \(m\) which is sufficiently large that the BTTN manifold of states is a good approximation for the full state space.
Because the single-center TDVP algorithm will not alter the bond dimension of this initial BTTN state, it is important that the initial state is constructed with careful control over the value of \(m\).
The DMRG algorithm~\cite{Silvi2019,schollwock2011density}, adapted to the BTTN framework and implemented using the ITensor library~\cite{itensor}, meets both needs.

Recall that one period of the flashing ratchet first acts with \(\mathsf{W}_1\) for time \(\tau / 2\) then with \(\mathsf{W}_2\) for time \(\tau / 2\).
In the large \(\tau\), slow switching limit, the time-periodic steady state at the end of a full period will be very similar to the time-independent steady state of \(e^{\mathsf{W}_2 \tau / 2}\), which of course shares eigenstates with the simpler \(\mathsf{W}_2\).
As a seed for TDVP, we therefore construct the top eigenstate of \(\mathsf{W}_2\), \(\left|\pi_2\right>\).
Because \(\mathsf{W_2}\) is a rate matrix, \(\left|\pi_2\right>\) has an associated eigenvalue of zero and has the physical interpretation of the (equilibrium) steady state for the zero-potential off state of the ratchet.
Furthermore, \(\mathsf{W_2}\) is Hermitian since it corresponds to a symmetric flat landscape.
We build a BTTN approximation to \(\left|\pi_2\right>\) by applying DMRG to \(\mathsf{W}_2\); convergence of the method is readily confirmed by comparing the obtained eigenvalue \(\psi_2\) with zero.

For a lattice with \(N\) sites, that DMRG algorithm is seeded with any pure state (a state in which a single amplitude \(c_{s_1\cdots s_N}\) in Eq.~\eqref{eq:productstate} is unity and the rest are zero) with exactly \(N_{\rm occ}\) occupied sites.
The occupancy of each site specifies the physical indices of that pure state, while the auxiliary indices are initially trivial with bond dimension 1. 
Using a block-sparse representation of the tensors~\cite{Silvi2019}, the number of particles is conserved so the resulting steady-state  \(\left|\pi_2\right>\) will be built only from states containing exactly \(N_{\rm occ}\) particles. 
To allow the bond dimension to grow and reach the targeted value \(m\), we implement single-site DMRG with subspace expansion~\cite{hubig2015strictly} with mixing parameter \(\alpha\).
The role of \(\alpha\) is to control the extent of the perturbative contribution from the expansion terms; too small a value could lead the perturbation terms becoming negligible, whereas too large a value could adversely interfere with DMRG convergence~\cite{hubig2015strictly,Yang2020a}.

For lattices with 128 or fewer sites, DMRG fully converges to the steady state of \(\mathsf{W_2}\) within a few dozen DMRG sweeps, though the convergence generally requires a sufficiently large \(m\) and a tuned value of \(\alpha\) which is neither too small nor too large~\cite{Yang2020a}.
It is important to realize that DMRG has more difficulty converging to the steady state as additional particles are added to the lattice.
Fig.~\ref{dmrgconv} shows these convergence trends for a 32-site lattice with \(N_{\rm occ} = 8\) and \(N_{\rm occ} = 16\).

\begin{figure*}[htb]
    \centering
    \begin{tikzpicture}
        \node at (-1,6.4) {\includegraphics[width=0.45\textwidth]{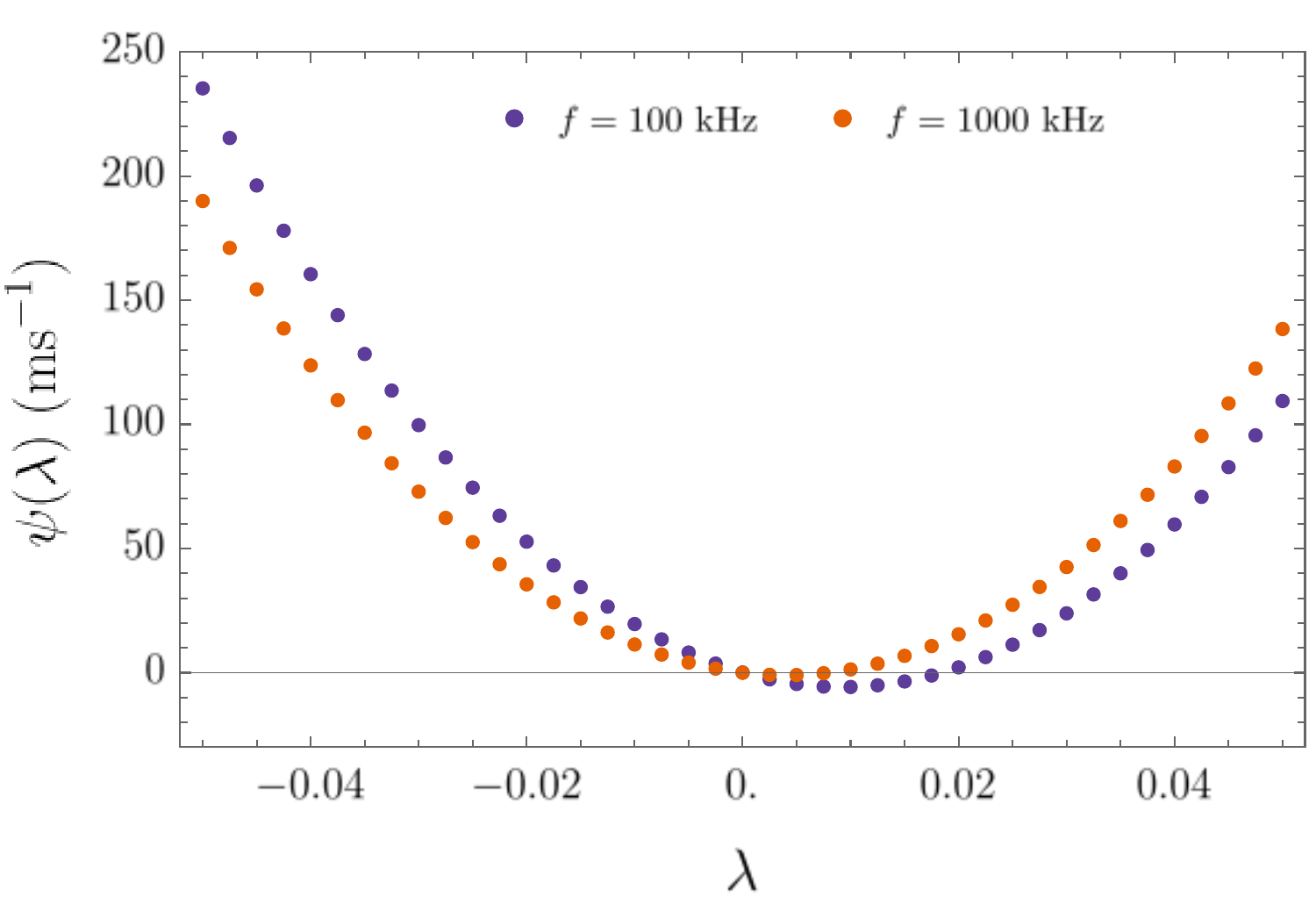}};
        \node at (8.5,6.4) {\includegraphics[width=0.45\textwidth]{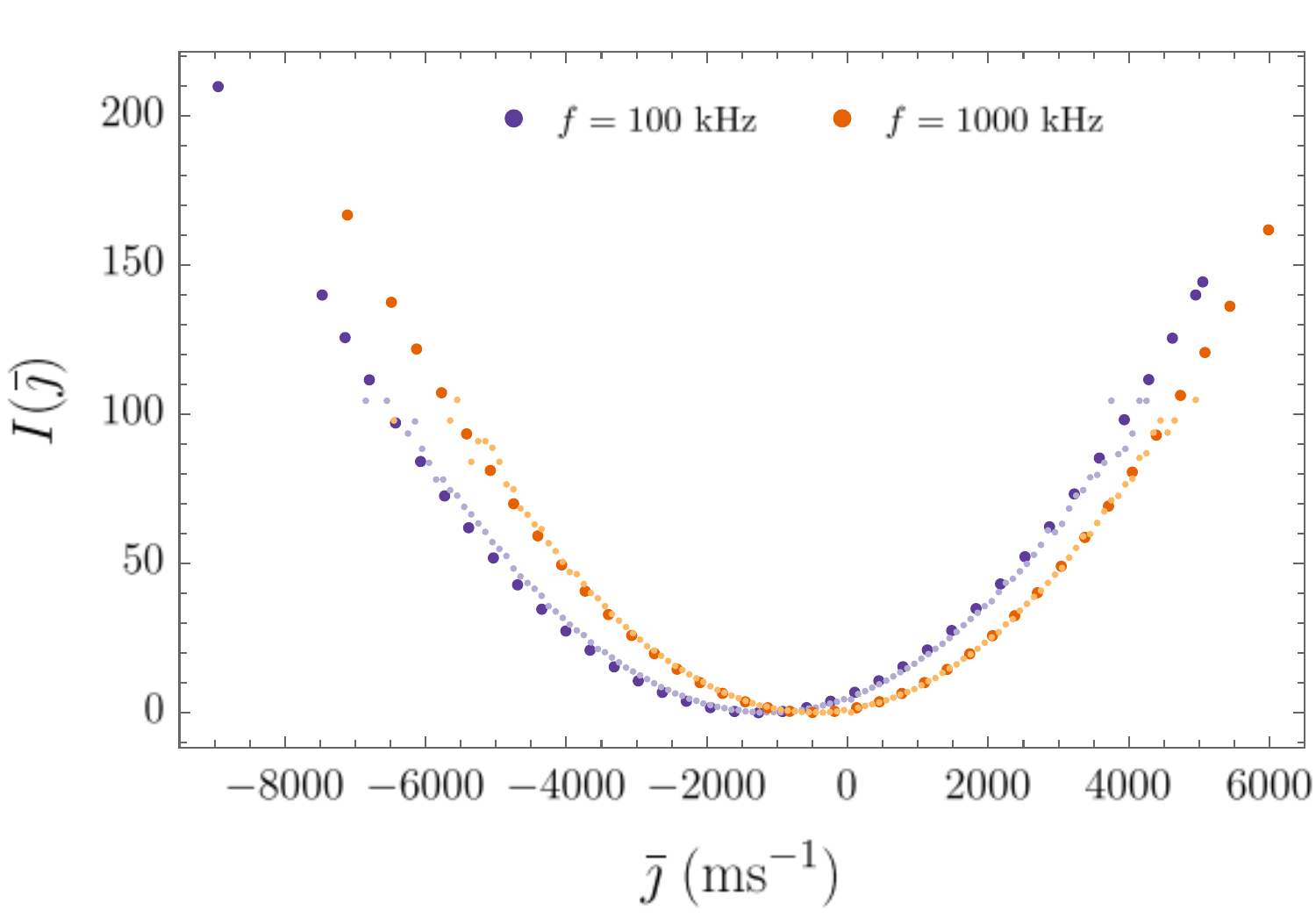}};
    \end{tikzpicture}
    \caption{Scaled cumulant-generating function (SCGF) and rate function obtained from TDVP.
    \textit{Left.} The SCGF \(\psi(\lambda)\) is plotted as a function of the biasing parameter \(\lambda\) for a 32-site lattice occupied by 16 particles and two driving frequencies, 100 and 1000~\si{\kHz}.
    The slope of the SCGF at \(\lambda=0\) is observed to have a greater magnitude when \(f=100~\si{\kHz}\) than when \(f=1000~\si{\kHz}\), in agreement with the trends in the currents seen in Fig.~\ref{jvm} as well as the companion letter~\cite{strand2021companion}.
    \textit{Right.} The SCGF values were used to compute rate functions \(I(\bar{\jmath})\), shown in thick dots, via a numerical Legendre transform.
    These rate functions were shown to be in very good agreement with Gillespie sampling (using the algorithm from Appendix~\ref{sec:gillespie}) of \(10^6\) \(t_{\rm obs} = 0.1~\si{\ms}\) long trajectories.
    Those Gillespie-sampled rate functions (small dots) fit a rate function from the histogram for \(P(\bar{\jmath})\) using \(I(\bar{\jmath}) = -\ln (P(\bar{\jmath}) - P(\left<\bar{\jmath}\right>)) / t_{\rm obs}\).
    }
    \label{ldt}
\end{figure*}

\subsection{Extracting the SCGF for currents from TDVP}
\begin{figure}[htb]
    \centering
    \begin{tikzpicture}
        \node at (0,6) {\includegraphics[width=0.45\textwidth]{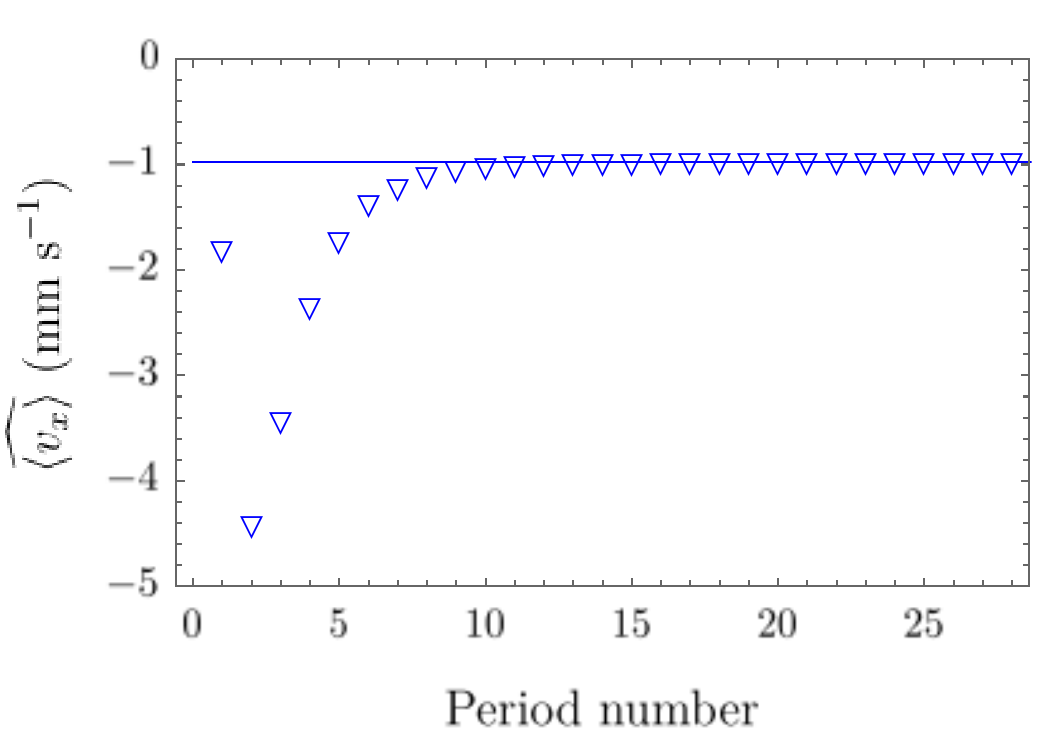}};
        \node at (0,0) {\includegraphics[width=0.45\textwidth]{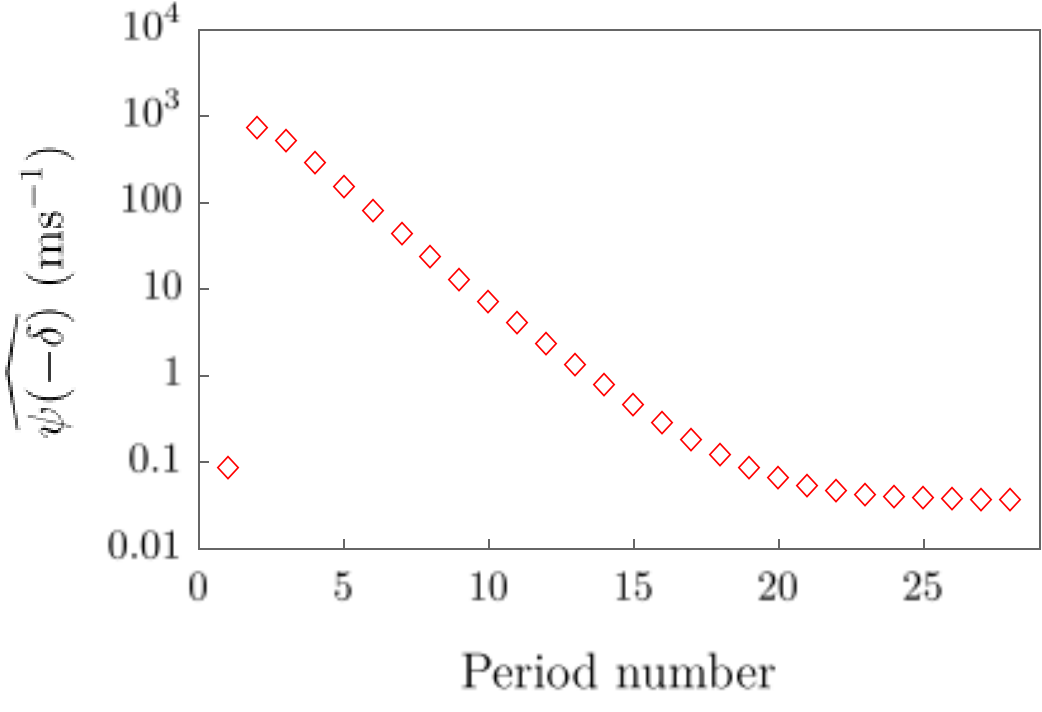}};
    \end{tikzpicture}
    \caption{Estimates (denoted with a hat) of \(\langle v_x\rangle\) and the SCGF \(\psi(\lambda)\) under weak biasing with \(\lambda = -\delta \equiv 10^{-4}\) are plotted for each period of TDVP evolution with timestep 1~\si{ns}.
    Results are shown for a 32-site lattice occupied by 16 particles and a driving frequency of 1~MHz.
    The corresponding value of the average particle velocity obtained from Gillespie sampling is represented by the blue horizontal line, whose thickness is 3 times the standard error.
    The mean current extracted from TDVP agrees with the Gillespie sampling in fewer periods of driving than are required to converge the SCGF.
    }
    \label{tdvpconv}
\end{figure}

The scaled cumulant-generating function (SCGF) \(\psi(\lambda)\) for period-averaged currents is computed via the TDVP evolution and Eq.~\eqref{eq:evolution}.
The resulting SCGF, plotted in Fig.~\ref{ldt}, contains information about the mean, variance, and higher cumulants of \(\bar{\jmath}\).
These statistical properties can be extracted from the behavior of the SCGF in the neighborhood of the origin, with the \(k^{\rm th}\) cumulant of \(\bar{\jmath}\) computed from the \(k^{\rm th}\) derivative of \(\psi(\lambda)\) evaluated at \(\lambda =0\).
Our companion letter focuses on mean currents, in which case we needed only the slope of the SCGF at \(\lambda = 0\)~\cite{strand2021companion}.
In practice, we compute the first derivative numerically by introducing a very small biasing strength \(\delta = 10^{-4}\) and approximating \(\left<\bar{\jmath}\right> = \psi'(0) \approx (\psi(\delta) - \psi(-\delta)) / (2 \delta)\).
Starting with the \(\lambda = 0\) seed \(\left|\pi_2\right>\), TDVP is run with a timestep \(\Delta t\) for enough periods to converge the mean steady-state current.
The calculation is stopped once the change in the current estimate between two adjacent periods lies within one percent of its magnitude, at which point full convergence is assumed.
Fig.~\ref{tdvpconv} illustrates the convergence of both \(\left<v_x\right>\equiv\left<\bar{\jmath}\right>h/N_\text{occ}\) and \(\psi(-\delta)\) over 20 periods of BTTN TDVP ratchet evolution with a half-occupied 32-site lattice.
Both quantities converge in the long-time limit, but the current's convergence is noticeably faster than that of the SCGF.

The rate of convergence depends on the frequency of driving, particularly because the DMRG-generated seed \(\left|\pi_2\right>\) is constructed to match the low-frequency limit.
For that reason, the low-frequency current can converge within one or two periods of TDVP evolution.
At high frequencies, it is necessary to run tens or hundreds of periods to allow \(\left|\pi_2\right>\) time to evolve into the time-periodic steady state.
One could converge more quickly by instead seeding with a high-frequency-limit eigenvector, the steady state of \((\mathsf{W_1}+\mathsf{W_2})/2\), but we found it sufficient (and simpler) to use the one seed for all frequencies.

The TDVP methodology extends beyond the small-\(\lambda\) regime, granting access also to fluctuations of \(\bar{\jmath}\).
These current fluctuations are characterized by a large deviation rate function \(I(\bar{\jmath})\), computed as a Legendre transform of \(\psi(\lambda)\)~\cite{touchette2009large}.
As a result, Fig.~\ref{ldt} shows that one can compute the distribution for the current averaged over \(n\) periods of driving, \(P(\bar{\jmath}) \simeq e^{-n I(\bar{\jmath})}\), by first performing TDVP tensor network calculations of \(\psi(\lambda)\) for various strengths of biasing \(\lambda\).
As a practical matter, those TDVP calculations are most stable if seeded by a state that approximates the steady-state \(\left|\pi(\lambda)\right>_t\).
We start by performing \(\lambda = 0\) calculations then increase and decrease \(\lambda\) in steps, seeding each calculation by a converged steady-state for a nearby value of \(\lambda\).

\begin{figure*}[htb]
    \centering
    \begin{tikzpicture}
        \node at (-1,6) {\includegraphics[width=0.45\textwidth]{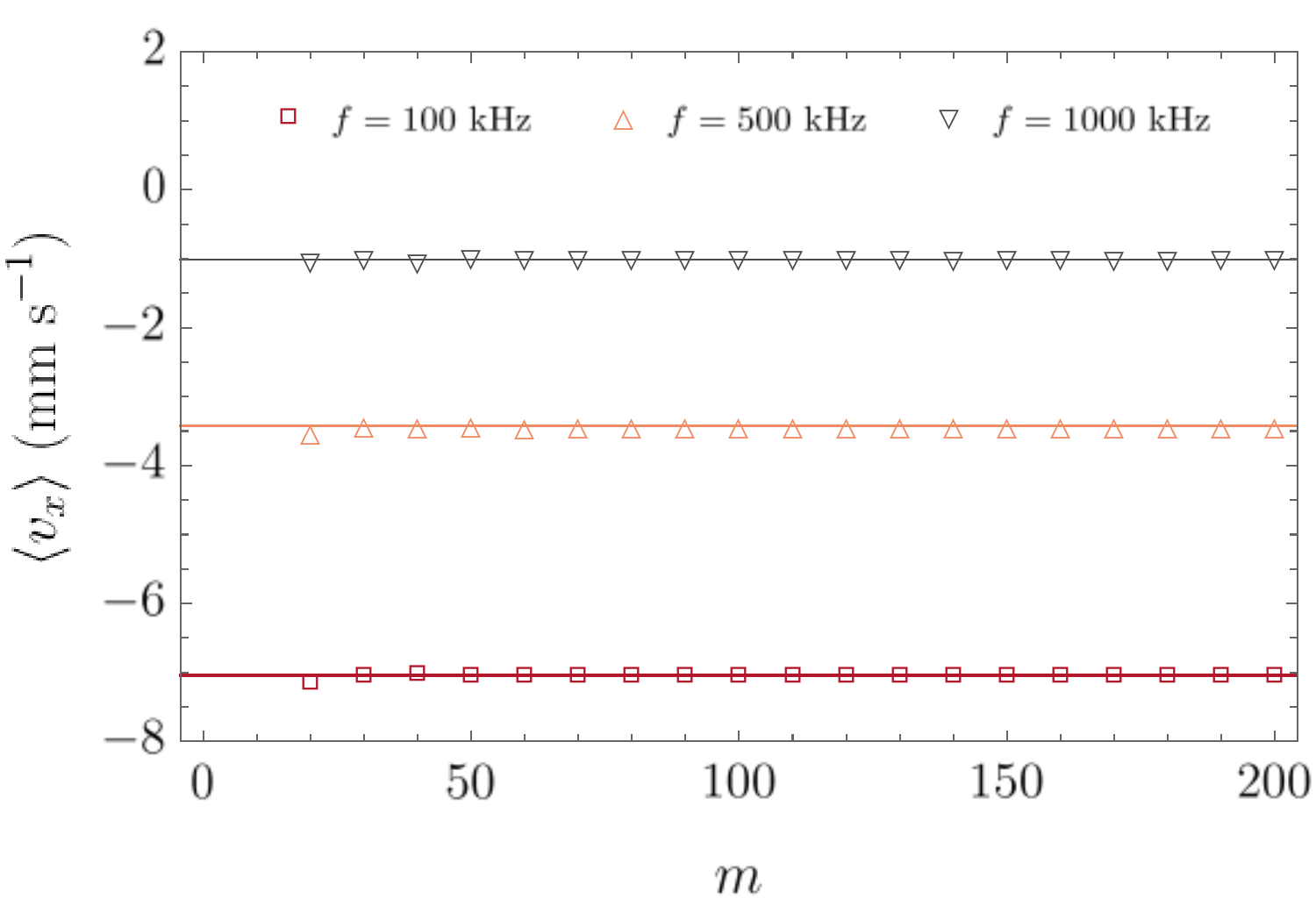}};
        \node at (8.5,6) {\includegraphics[width=0.45\textwidth]{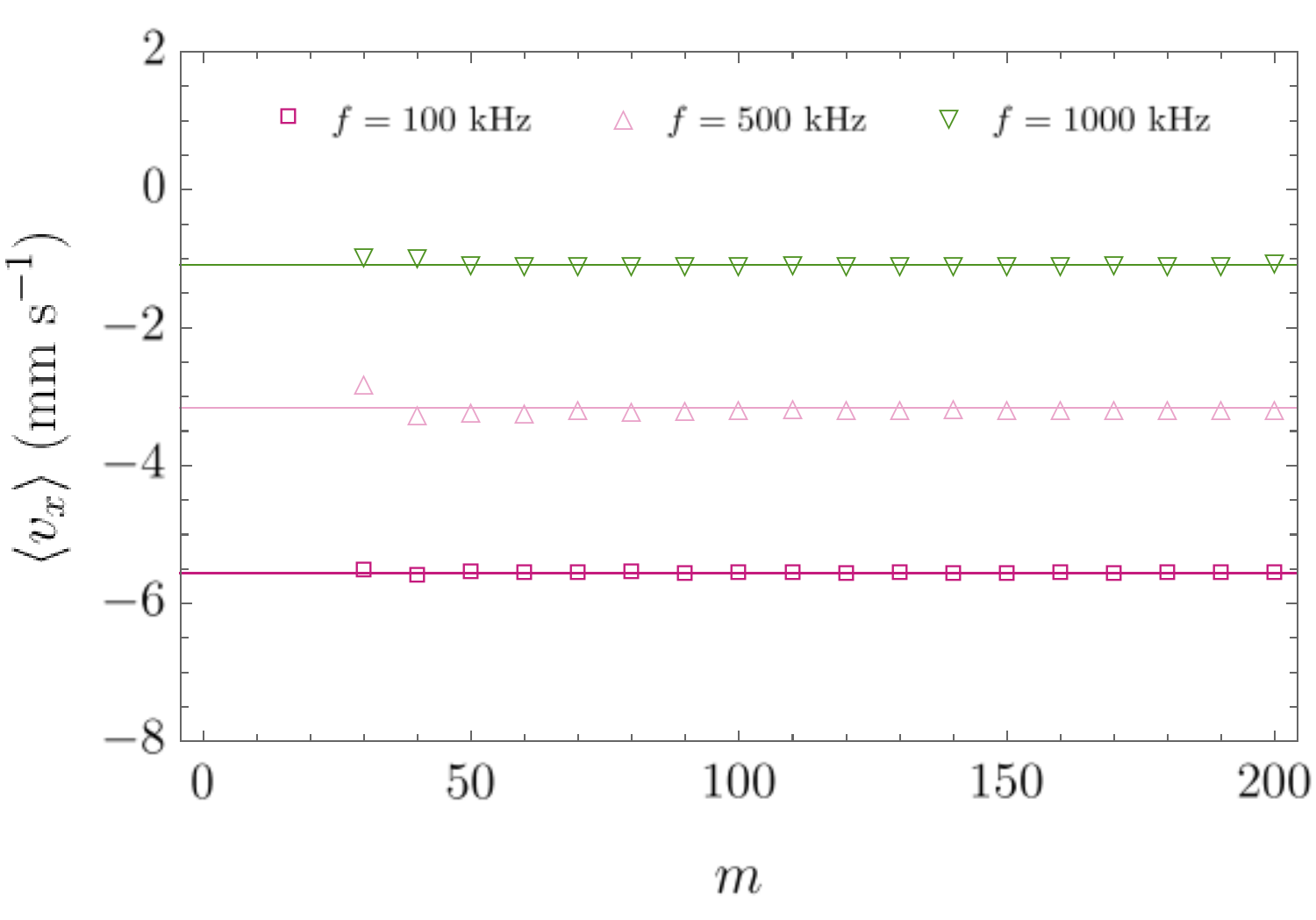}};
        \node at (-1,0) {\includegraphics[width=0.45\textwidth]{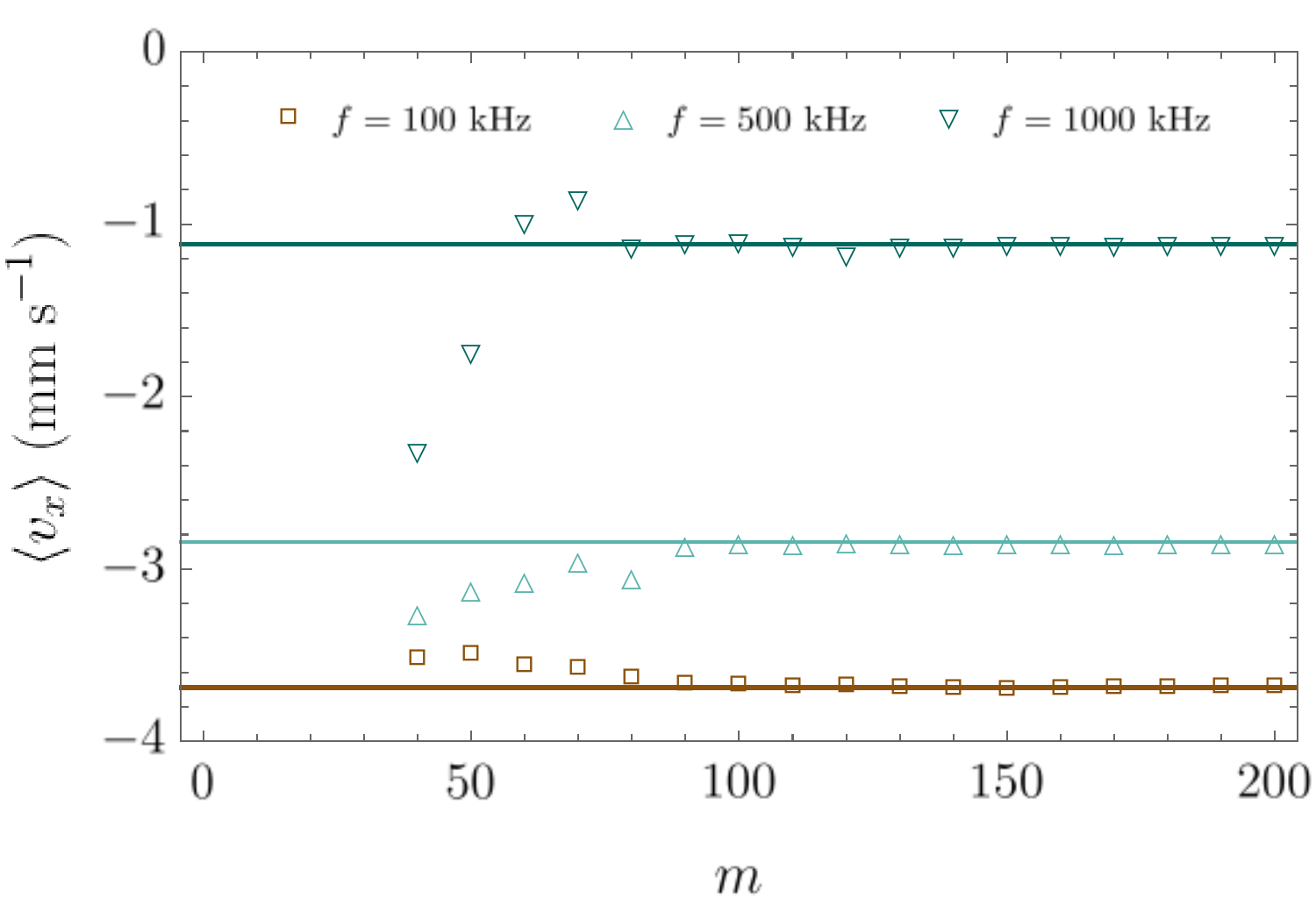}};
        \node at (8.5,0) {\includegraphics[width=0.45\textwidth]{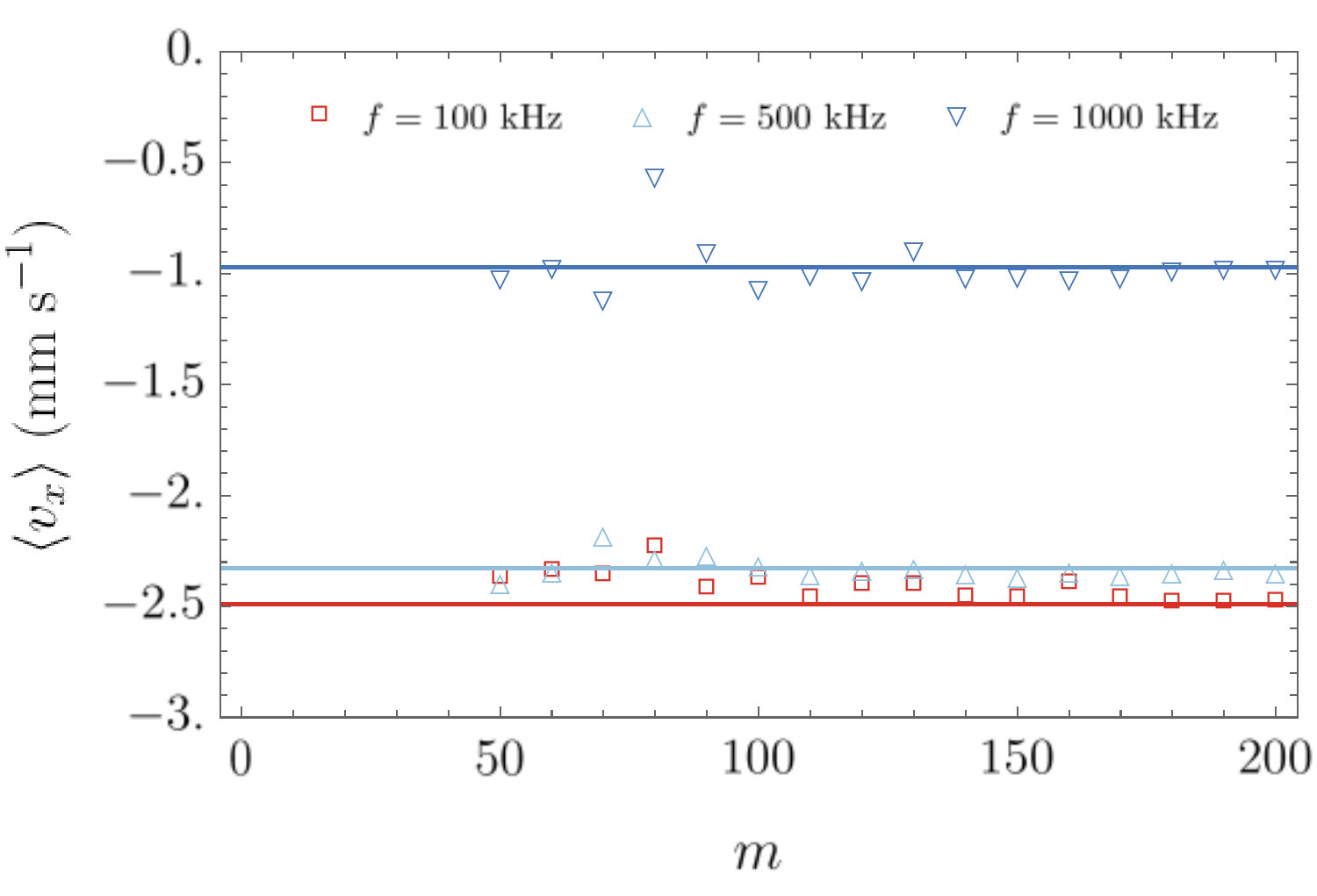}};
        \node at (-4.5,8) {(a)};
        \node at (5,8) {(b)};
        \node at (-4.5,2) {(c)};
        \node at (5,2) {(d)};
    \end{tikzpicture}
    \caption{Period-averaged mean particle velocity computed from TDVP as a function of the maximal bond dimension.
    Average particle velocities for a 32-site lattice are plotted against \(m\) for \(N_{\rm occ} = 4\) (a), 8 (b), 12 (c), and 16 (d) with driving frequencies \(f = 100, 500,\) and \(1000~\si{kHz}\).
    TDVP calculations with \(N_{\rm occ} = 4, 8, and 12\) used \(\Delta t = 1~\si{ns}\).
    The \(N_{\rm occ} = 16\) calculations used \(\Delta t = 0.1~\si{ns}\) to mitigate numerical instabilities that were especially prominent for \(m < 150\).
    DMRG fails to converges when \(m\) is very small, particularly for high occupancy.
    For those small \(m\) values, the TDVP calculation was not performed because it could not be seeded by \(\left|\pi_2\right>\).
    Average particle velocity obtained from Gillespie sampling are represented by horizontal lines, whose thicknesses are 3 standard errors.
    As the number of particles occupying the lattice increases, the required \(m\) increases.}
    \label{jvm}
\end{figure*}

\subsection{Comparison with Monte Carlo sampling}
As a variational method, TDVP is not assured to work for small \(m\).
We validate that the tensor network ansatz indeed provides a good approximation by comparing with kinetic Monte Carlo sampling of the discrete-state jump process via Gillespie sampling.
Due to switches between \(\mathsf{W_1}\) and \(\mathsf{W_2}\), however, waiting times for a hop no longer come from an exponential distribution and the usual Gillespie algorithm be modified.
Anderson~\cite{anderson2007modified} has developed a rejection-based stochastic simulation algorithm (RSSA) to handle Markovian jump processes with arbitrary time dependencies.
Because our 1D system relies on a square wave driving protocol, the usual Gillespie algorithm can be modified more simply.
We describe the specific algorithm in Appendix~\ref{sec:gillespie}.
To estimate mean currents, 512 independent Gillespie trajectories were averaged.
Each trajectory was allowed to relax to its time-periodic steady state by a 0.01~\si{ms} burn-in followed by a measurement of the current generated in 100~\si{ms}.

Those Gillespie calculations of mean currents are simpler and less expensive than the tensor network methodology, but the TDVP approach offers some unique benefits.
Fig.~\ref{ldt} illustrates that the TDVP calculations accurately predict rare current fluctuations, even fluctuations that are more rare than can be readily observed by straightforward unbiased Gillespie sampling.
Furthermore, the TDVP approach naturally generalizes to \(\mathsf{W}(t)\) with arbitrary  time-dependence whereas our Gillespie approach of Appendix~\ref{sec:gillespie} is specialized to the square-wave temporal driving.
More general time dependence would require a more costly Gillespie strategy like RSSA.

\subsection{TDVP with varied bond dimension}
\label{sec:mainresults}

The computational expense of the TDVP grows rapidly with the maximum bond dimension \(m\).
Consequently, to practically compute steady state properties from the TDVP, it is essential that the \(m\) can be kept small while maintaining accuracy.
We numerically probed the needed bond dimension by repeating the TDVP calculations on an \(N = 32\) lattice with a range of \(m\), adjusted via the DMRG seed \(\left|\pi_2\right>\).
These calculations were carried out for a range of driving frequencies and \(N_{\rm occ}\) values.
An optimal TDVP timestep depends on both \(N\) and \(N_{\rm occ}\).
Too large a timestep results in numerical instability and convergence issues; too small makes a calculation unnecessarily costly.
To compare the bond dimension results most simply, we used a fixed timestep of \(\Delta t =1~\si{ns}\), except for the case of \(N_{\rm occ} = 16\) which required \(\Delta t = 0.1~\si{ns}\) to accurately converge.

The dependencies on maximum bond dimension are shown in Fig.~\ref{jvm}.
In all cases, the TDVP current tends to the value obtained from Gillespie simulation with a large enough \(m\).
This bond dimension threshold increases the more particles occupy the lattice, as rationalized by the vast increase in the number of states accessible by TDVP as particles are added to lattice.
When the lattice is occupied by only 4 particles, a maximal bond dimension of merely 30 is sufficient for TDVP to produce accurate ratchet currents within the driving frequency range considered, whereas the required maximal bond dimension increases dramatically (to around 180) for a half-occupied lattice (\(N_\text{occ}=16\)).

\section{Discussion}
\label{sec:discussion}

We have illustrated that a BTTN with a tractable maximum bond dimension is sufficient to propagate a distribution over many-particle states evolving under a time-periodic protocol.
The more conventional Gillespie approach evolves a single trajectory at a time, then average over those trajectories.
Propagating the distribution via TDVP complements that strategy and offers several potential benefits.
Firstly, as shown in Fig.~\ref{ldt}, the TDVP approach naturally gives information about both typical and rare events \emph{at comparable computational expense}.
While Gillespie sampling can also be biased to probe rare events, those calculations typically require significantly more computational power than the unbiased sampling of typical events.
Secondly, the TDVP approach naturally generalizes to time-dependent rate matrix, a situation that can be quite challenging for Gillespie sampling.
Finally, our calculations have repeated dynamics for different systems parameters, for example different frequencies \(f\).
In the case of Gillespie sampling, the change in parameters demands an entirely new batch of simulated trajectories.
The prior calculations do not speed up the next batch, which has to be sampled from scratch.
As calculation on the whole distribution, the TDVP calculations can leverage prior calculations to more rapidly converge steady-state dynamics with similar system parameters.
We wrote about seeding our TDVP evolution from the state \(\left|\pi_2\right>\), but it can also be seeded from the converged state reached by a prior calculation.
For example, suppose one needs to compute \(\psi(\lambda)\) for various frequencies.
The converged calculation with frequency \(f_1\) and biasing strength \(\lambda\) will have settled into a time-periodic state \(\left|\pi(\lambda, f_1)\right>_t\), which can be the initial state for the TDVP dynamics used to estimate \(\psi(\lambda)\) at frequency \(f_2\).
Depending on the application, we anticipate that this ability to leverage prior calculations could warrant the extra complexity of the tensor network approach.

The present work is a first attempt to employ tensor networks to treat time-periodic steady states in many-particle classical stochastic dynamics.
Given the exceptional advances in tensor network methodologies, we anticipate future improvements to the stability and efficiency of the types of calculations we have described.
Efficient new ways to compute time-evolution operators~\cite{vanhecke2021simulating}, adaptive timesteps, and algorithms that adaptively construct tree tensor networks based on the structure of the rate matrix~\cite{ferrari2021adaptive} could all offer a path to future optimizations and improvements.

\section{Acknowledgments}
We gratefully acknowledge Schuyler Nicholson and Phillip Helms for many insightful discussions.
We are also grateful to Miles Stoudenmire, Matthew Fishman, Steven White, and other developers of ITensor, a library for implementing tensor network calculations, upon which this work was built.
The material presented in this manuscript is based upon work supported by the National Science Foundation under Grant No.\ 2141385.

\appendix
\section{Matrix product operator (MPO) representation of tilted operators}
\label{sec:mpo}
Eq.~\eqref{eq:secondquant} gives a compact representation of \(\mathsf W_k(\lambda)\) that sums over all nearest-neighbor pairs of sites around the periodic boundary conditions.
It is convenient, however, to deconstruct that sum in terms of a product of operator-valued vectors and matrices.
The decomposition can be performed identically for each \(k\).
For compactness, we suppress the subscript \(k\) and write that matrix product as
\begin{equation}
  \mathsf W(\lambda) = W(1) W(2) \cdots W(N),
\end{equation}
where \(W(1)\) is a one-by-ten row vector, \(W(N)\) is a ten-by-one column vector, and the other \(W(i)\)'s are ten-by-ten matrices.
By factorizing Eq.~\eqref{eq:secondquant} in this manner, the tilted rate matrix is seen to be an MPO with each \(W(i)\) corresponding to a shaded gray circle in Fig.~\ref{diagram}a.
The ITensor library~\cite{itensor} contains an AutoMPO function that factorizes a sum like Eq.~\eqref{eq:secondquant} into an explicit MPO.
Alternatively, a finite-state machine can be employed~\cite{schollwock2011density} to derive the factorized local tensors for sites 1, \(i = 2, \hdots N-1\), and N, that are given by
\begin{widetext}
\begin{align}
\nonumber  W(1)&=\begin{pmatrix}
  \mathbf 0 & r_{1\to2}e^\lambda\mathbf a & -r_{1\to2}\mathbf n & -r_{2\to1}\mathbf v & r_{2\to1}e^{-\lambda}\mathbf a^\dagger & r_{1\to N}e^{-\lambda}\mathbf a& -r_{1\to N}\mathbf n & -r_{N\to1}\mathbf v & r_{N\to1}e^\lambda\mathbf a^\dagger & \mathds I
  \end{pmatrix},\\
\nonumber  W(i)&=\begin{pmatrix}
  \mathds I & \mathbf 0 & \mathbf 0 & \mathbf 0 & \mathbf 0 & \mathbf 0 & \mathbf 0 & \mathbf 0 & \mathbf 0 & \mathbf 0\\
  \mathbf a^\dagger & \mathbf 0 & \mathbf 0 & \mathbf 0 & \mathbf 0 & \mathbf 0 & \mathbf 0 & \mathbf 0 & \mathbf 0 & \mathbf 0\\
  \mathbf v & \mathbf 0 & \mathbf 0 & \mathbf 0 & \mathbf 0 & \mathbf 0 & \mathbf 0 & \mathbf 0 & \mathbf 0 & \mathbf 0\\
  \mathbf n & \mathbf 0 & \mathbf 0 & \mathbf 0 & \mathbf 0 & \mathbf 0 & \mathbf 0 & \mathbf 0 & \mathbf 0 & \mathbf 0\\
  \mathbf a & \mathbf 0 & \mathbf 0 & \mathbf 0 & \mathbf 0 & \mathbf 0 & \mathbf 0 & \mathbf 0 & \mathbf 0 & \mathbf 0\\
  \mathbf 0 & \mathbf 0 & \mathbf 0 & \mathbf 0 & \mathbf 0 & \mathds I & \mathbf 0 & \mathbf 0 & \mathbf 0 & \mathbf 0\\
  \mathbf 0 & \mathbf 0 & \mathbf 0 & \mathbf 0 & \mathbf 0 & \mathbf 0 & \mathds I & \mathbf 0 & \mathbf 0 & \mathbf 0\\
  \mathbf 0 & \mathbf 0 & \mathbf 0 & \mathbf 0 & \mathbf 0 & \mathbf 0 & \mathbf 0 & \mathds I & \mathbf 0 & \mathbf 0\\
  \mathbf 0 & \mathbf 0 & \mathbf 0 & \mathbf 0 & \mathbf 0 & \mathbf 0 & \mathbf 0 & \mathbf 0 & \mathds I & \mathbf 0\\
  \mathbf 0 & r_{i\to i+1}e^\lambda\mathbf a & -r_{i\to i+1}\mathbf n & -r_{i+1\to i}\mathbf v & r_{i+1\to i}e^{-\lambda}\mathbf a^\dagger & \mathbf 0 & \mathbf 0 & \mathbf 0 & \mathbf 0 & \mathds I\\
  \end{pmatrix},\\
  W(N)&=\begin{pmatrix}
  \mathds I & \mathbf a^\dagger & \mathbf v & \mathbf n & \mathbf a & \mathbf a^\dagger & \mathbf v & \mathbf n & \mathbf a & \mathbf 0
  \end{pmatrix}^{\mathsf T}.
\end{align}
\end{widetext}

\section{Gillespie algorithm for square wave driving}
\label{sec:gillespie}
Because the flashing ratchet has a temporal drive with a period of \(\tau\), standard Gillespie sampling~\cite{gillespie1977exact} must be adapted to accommodate the time-dependent propensities.
As in the traditional algorithm, these propensities are used to compute a target state as well as a random waiting time at each step along a trajectory.
Where the traditional algorithm breaks down is in the event that a drawn waiting time would span both sets of propensities.
For example, if the previous hop occurred during a \(\mathsf{W_1}\) propagation but the next would not occur until the \(\mathsf{W_2}\) propagation, then the waiting time should reflect some mixture of the \(\mathsf{W_1}\) and \(\mathsf{W_2}\) rates.

Letting the time-dependent propensity be denoted by \(w(t)\), the waiting time \(\delta t\) should solve
\begin{equation}
    \int_{t_0}^{t_0+\delta t}w(t)dt=\ln\left(\frac{1}{s}\right),
    \label{eq:propint}
\end{equation}
where \(s \sim U(0,1)\) is a random number drawn uniformly from the unit interval and \(t_0\) is the time of the most recent hop~\cite{anderson2007modified}.
  When \(w(t)\) is a constant \(w\), the integral evaluates to \(w \delta t\), recovering the usual Gillespie algorithm for drawing waiting times.
  While it is not as simple, the integral can be similarly evaluated for the square wave driving that flips between a rate \(w_1\) and another rate \(w_2\).

  Without loss of generality, let us assume \(w(t_0) = w_1\).
  If \(w(t)\) remains \(w_1\) until \(t=t_0+\delta t\), that is, if \(\text{Mod}(t_0,\tau)+\delta t<\tau/2\), then \(\delta t\) is computed as usual, namely
  \begin{equation}
    \delta t=(1/w_1)\ln(1/s).
    \label{eq:case1}
  \end{equation}
  If the waiting time \(\delta t\) would pass through the time that the rate jumps from \(w_1\) to \(w_2\), then Eq.~\eqref{eq:propint} instead integrates to give
  \begin{equation}
    \left(\frac{\tau}{2}-t_0\right)w_1+\left(t_0+\delta t-\frac{\tau}{2}\right)w_2=\ln\left(\frac{1}{s}\right),
\end{equation}
which, after some algebra, yields
\begin{equation}
  \delta t=\frac{1}{w_2}\left[\ln\left(\frac{1}{s}\right)+(w_2-w_1)\left(\frac{\tau}{2}-t_0\right)\right].
  \label{eq:case2}
\end{equation}
Particularly when the driving frequency is high, it is possible that a waiting time \(\delta t\) could pass through the time that \(w_1\) switches to \(w_2\) as well as the time that the rate switches back to \(w_1\).
In that case, the waiting time is related to the random number \(s\) as
\begin{equation}
  \left(\frac{\tau}{2}-t_0\right)w_1+\frac{\tau}{2}w_2+\left(t_0+\delta t-\tau\right)w_1=\ln\left(\frac{1}{s}\right),
\end{equation}
leading to the waiting time
\begin{equation}
  \delta t=\frac{1}{w_1}\left[\ln\left(\frac{1}{s}\right)+(w_1-w_2)\frac{\tau}{2}\right].
\end{equation}
One can continue casing out the possibilities, adding more cycles between \(w_1\) and \(w_2\) before the next hop occurs.
For example, the next case involves waiting time
\begin{equation}
    \delta t=\frac{1}{w_2}\left[\ln\left(\frac{1}{s}\right)+(w_2-w_1)(\tau-t_0)\right].
\end{equation}

In practice, one starts by computing the rates \(w_1\) and \(w_2\) associated with each possible jump from the current configuration at time \(t_0\).
Next, \(s\) is drawn and a separate waiting time \(\delta t\) is computed for each possible hop.
\(w(t_0)\) is set to \(w_1\) if \(\text{Mod}(t_0,\tau)<\tau/2\), and to \(w_2\) otherwise.
For a given hop, if Eq.~\eqref{eq:case1} yields a \(\delta t\) consistent with the first case, meaning a \(\delta t\) sufficiently short that the square wave drive will not have switched from \(\mathsf{W_1}\) to \(\mathsf{W_2}\) (or from \(\mathsf{W_2}\) to \(\mathsf{W_1}\)), then that waiting time is chosen.
Otherwise, we proceed to the next case, inserting that \(s\) into Eq.~\eqref{eq:case2} (if \(w(t_0)=w_1\)).
We continue passing through the cases until the computed \(\delta t\) is consistent with the considered case for how many square wave flips have been experienced.
Once consistency is achieved, that \(\delta t\) is taken to be the next waiting time for that proposed transition.
Finally, the next chosen transition is the one with the smallest waiting time.
%All that remains is to take a hop to a new state at that time \(t_0 + \delta t\).
%This hop proceeds as in usual Gillespie, but with the rates given by \(\mathsf{W}(t_0 + \delta t)\), that is to say the rates at the instant of the hop~\cite{anderson2007modified}.

\bibliography{biblio.bib}

\end{document}